\documentclass[a4paper, 11pt]{article}
\usepackage[utf8]{inputenc}
\usepackage[backend=biber,style=numeric]{biblatex}
\usepackage{amsmath}
\usepackage[normalem]{ulem}
\usepackage[margin=1in]{geometry}

\usepackage{mathrsfs}
\usepackage{algorithm}
\usepackage{algorithmic}
\usepackage{amssymb}
\usepackage{amsmath}
\usepackage{amsthm}
\usepackage{amsfonts}
\usepackage{enumitem} 
\usepackage{dsfont}
\usepackage{booktabs}

\usepackage[colorlinks,unicode]{hyperref}
\usepackage[nameinlink,capitalise,noabbrev]{cleveref}

\newcommand*{\gs}[1]{{\color{red}GS: #1}}
\newcommand*{\ev}[1]{{\color{orange}[#1]}}
\newcommand*{\af}[1]{{\color{blue}[#1]}}
\newcommand*{\nitya}[1]{{\color{teal}[#1]}}
\newcommand*{\resolved}[1]{}

\allowdisplaybreaks

\makeatletter
\DeclareRobustCommand*{\bfseries}{
  \not@math@alphabet\bfseries\mathbf
  \fontseries\bfdefault\selectfont
  \boldmath
}
\makeatother

\theoremstyle{plain}
\newtheorem{theo}{Theorem}
\newtheorem{conj}[theo]{Assumption}
\newtheorem{prop}[theo]{Proposition}
\newtheorem{lem}[theo]{Lemma}
\newtheorem{cor}[theo]{Corollary}
\theoremstyle{definition}
\newtheorem{defn}[theo]{Definition}
\theoremstyle{remark}
\newtheorem{rem}[theo]{Remark}

\DeclareUnicodeCharacter{2212}{-}
\DeclareUnicodeCharacter{FB01}{-}
\DeclareUnicodeCharacter{FB03}{-}
\usepackage{graphicx}
\usepackage{float}
\graphicspath{{figures/}}
\usepackage[labelfont=bf]{caption}
\usepackage{subcaption}

\usepackage[toc,page]{appendix}

\usepackage{tikz}
\usetikzlibrary{positioning, decorations, arrows, shapes}
\tikzset{
    rndproc/.style={circle,draw, thick}
}
\tikzset{
    hidden/.style={rndproc, fill=white!20!gray}
}

\usepackage{hyperref}
\hypersetup{
    colorlinks=true,
    citecolor=red,
    linkcolor=blue,
    linktoc=all
}

\usepackage{titlesec}
\titleformat{\paragraph}
{\normalfont\normalsize\bfseries}{\theparagraph}{1em}{}
\titlespacing*{\paragraph}
{0pt}{3.25ex plus 1ex minus .2ex}{1.5ex plus .2ex}

\author{Elias Ventre}

\begin{document}

\begin{center}
\noindent \uppercase{\textbf{Trajectory inference for a branching SDE model of cell differentiation}}
\end{center}
    \vspace{.5cm}
\noindent \textbf{Authors}:
{Elias Ventre}$^{1}$,
{Aden Forrow}$^{2}$,
{Nitya Gadhiwala}$^{1}$,
{Parijat Chakraborty}$^{1}$,
{Omer Angel}$^{1}$,
{Geoffrey Schiebinger}$^{1}$.\\

\noindent 1 - {The University of British Columbia, Vancouver, BC, Canada}.\\
2 - {The University of Maine, Orono, Maine, USA}.\\
\noindent \textbf{Corresponding author}: {geoff@math.ubc.ca}.\par

\begin{abstract}
%A theory of trajectory inference has been recently developed for reconstructing, from time-courses of high-dimensional gene expression data, the landscape driving the differentiation of cells. The resulting method provides rigorous theoretical guarantees only when the birth and death of cells can be neglected. Otherwise, it requires in addition, like all other trajectory inference methods, the precise knowledge of the birth and death rates prior to the observation. Motivated by the recent development of CRISPR-based measurement technologies allowing to reconstruct the lineage tree of a population of cells, we show how such information can be used to tackle these limitations. In particular, when there is no death nor subsampling of cells, we show that we can extend the method to the case with proliferation with similar theoretical guarantees and computational cost, without requiring any prior information. In case of death and/or subsampling, this new trajectory inference method introduces a bias, that we describe explicitly and argue to be inherent to these lineage tracing data. We demonstrate in both cases the ability of this method to reliably reconstruct the time-varying distribution of a branching SDE from time-courses of simulated datasets with lineage tracing, even surpassing the results obtained previously when the true branching rates are known instead of lineage tracing.
A core challenge for modern biology is how to infer the trajectories of individual cells from population-level time courses of high-dimensional gene expression data. Birth and death of cells present a particular difficulty: existing trajectory inference methods cannot distinguish variability in net proliferation from cell differentiation dynamics, and hence require accurate prior knowledge of the proliferation rate. Building on Global Waddington-OT (gWOT), which performs trajectory inference with rigorous theoretical guarantees when birth and death can be neglected, we show how to use lineage trees available with recently developed CRISPR-based measurement technologies to disentangle proliferation and differentiation. In particular, when there is neither death nor subsampling of cells, we show that we extend gWOT to the case with proliferation with similar theoretical guarantees and computational cost, without requiring any prior information. In the case of death and/or subsampling, our method introduces a bias, that we describe explicitly and argue to be inherent to these lineage tracing data. We demonstrate in both cases the ability of this method to reliably reconstruct the landscape of a branching SDE from time-courses of simulated datasets with lineage tracing, outperforming even a benchmark using the experimentally unavailable true branching rates.
\end{abstract}

\vspace{.5cm}
\par

\bigskip


\tableofcontents
\newpage

\section{Introduction}

Over the last twenty years, single-cell RNA-sequencing (scRNA-seq) technologies have opened new windows on the study of cell-differentiation processes by offering the possibility of measuring, for a population of cells, the numbers of mRNA molecules expressed by the genes of each individual cell. In probabilistic terms, these \emph{single-cell data} give access to a joint probability distribution of gene expression, instead of the average value for each gene provided by population-level data~\cite{Mar2019,Coskun2016}. A major limitation is that the observation process is destructive: instead of observing the trajectory of cells directly, we can only observe the state, in the so-called gene expression space, of independent cells at different timepoints. Single-cell data can then be seen as temporal marginals of a stochastic process
characterizing the differentiation process. Trajectory inference is the challenging problem of recovering the full dynamic process from the limited static snapshots.

\bigskip

A large and rapidly growing set of methods have been proposed for inferring trajectories using a wide range of computational tools, including diffusion maps~\cite{Haghverdi2016}, recurrent neural networks~\cite{hashimoto2016learning}, and kernel regression~\cite{Qiu2022}. The algorithms are commonly based on heuristic biological intuition and rarely come with rigorous theoretical guarantees. One exception is the optimal transport based theory introduced with Waddington-OT (WOT)~\cite{Schiebinger2019}, which reconstructs the associated time-varying probabilistic distribution from time-courses of potentially sparse high-dimensional gene expression data. 
%\gs{The wording above can be improved. Maybe above should change to: "which connects ancestors to descendants with optimal transport"?}\ev{I would prefer not to, because it's not directly the notion of coupling between cells that provides theoretical guarantees}
Lavenant et al.~\cite{Lavenant2021} provided a careful mathematical foundation for a variant of WOT named global Waddington-OT (gWOT). In particular, under the hypothesis that the differentiation process can be modeled by a Stochastic Differential Equation (SDE) where the drift has zero curl (and hence is the gradient of some potential as in equation \eqref{eq_SDE} below), they proved that the reconstructed measure converges to the ground truth as the number of timepoints at which the cells are measured goes to infinity. 

%Note that many other methods of trajectory inference exist with the SDE model. However, they generally focus on recovering directly the potential of the SDE that they encode with a neural network architecture \cite{tong2020trajectorynet, hashimoto2016learning}, losing then the convexity of the problem which makes it difficult to establish rigorous theoretical guarantees.  \af{Mention other less rigorous trajectory inference techniques?}\ev{I've added this last sentence..}

\bigskip

However, the analysis by Lavenant et al does not account for cellular proliferation and death.  
%The SDE assumption is a limitation as it does not take into account the divisions and deaths of cells that are likely to play a major role in the observations: proliferation of cells is strongly related to their position in the gene expression space. 
% GSr: could delete the above and get to this next sentence quicker. 
Extending the theory to the case of \emph{branching processes} {is a major challenge when the proliferation and death of cells 
%, at exponential rates, 
allow for variations in mass that are nonuniform across the gene expression space}. In the theory of gWOT, the authors presented a way to adjust for proliferation if the ground truth growth rates were known before the experiment \cite{Lavenant2021, zhang2022trajectory}, but their method presented two major drawbacks: i) it was very sensitive to the prespecified proliferation rate and ii) convergence to the ground-truth path-measure was no longer guaranteed. As the characteristics of the SDE and the rates characterizing the branching process are strongly intertwined, these limitations seem difficult to overcome with sc-RNAseq data. To the best of our knowledge, all existing methods that attempt to account for branching require prior knowledge of the branching rates~\cite{weinreb2018fundamental, zhang2022trajectory}. 

\bigskip

In this paper, we show how recent {\em lineage tracing} technologies, which measure not only cell state but also the lineage relationships in a population of cells~\cite{mckenna2016whole, raj2018simultaneous} provide a solution to the growth rate problem in trajectory inference.
One approach to lineage tracing leverages CRISPR-Cas9 to continuously mutate an array of synthetic DNA barcodes that have been incorporated into the chromosomes so that they are inherited by daughter cells. By analyzing the pattern of mutations in the barcodes of each cell in a population, biologists can reconstruct a lineage tree which describes shared ancestry within the population. \resolved{\gs{We could say this is similar to the problem of phylogenetic inference.}\ev{Aden, as you worked on it, could you see how to integrate a sentence with a reference?}TO DO LATER} At a given time of measurement, this lineage tree then encodes both the position of cells in the gene expression space at its leaves, and for every pair of cells the time of their most recent common ancestor. However, it does not contain any direct information about the state of this ancestor. Thus, this technology makes accessible a new kind of data, for which the measurement of a population of cells can be seen as a set of \emph{trees} rather than a set of \emph{cells}.

\bigskip 

The main result of this article is that this additional lineage information can be used to {extend the theory of trajectory inference to branching processes.} %old version: which extends the theory of trajectory inference developed for probabilistic processes without requiring any additional prior information}. 
%GSr: shorten the above. 
In the particular case where the cells cannot die and all the leaves of each tree are well observed, we develop a method which is asymptotically exact (in the limit of an infinite number of timepoints at which the cells are collected, as in Lavenant et al.~\cite{Lavenant2021}). We also show that this method allows for an efficient estimation of the characteristics of the branching process, that is the drift of the underlying SDE and the division rates. Importantly, we demonstrate that our method even outperforms the result provided by the gWOT method with full prior knowledge of the branching rates~\cite{zhang2022trajectory}.%\gs{Wording to improve for the last sentence}\ev{Should be resolved}
In the general setting where the death rate is nonzero or not all cells are observed, a bias arises in the reconstructed path-measure due to the incompleteness of the information used in the lineage tree. We characterize this bias explicitly and propose a heuristic method to reduce it. Although removing the bias completely seem out of reach with this type of data, we argue and illustrate numerically that this bias is likely to be small, even without applying the heuristic correction.

\bigskip

We emphasize that these results are very different from the few works previously published in this nascent field of single-cell data analysis with lineage tracing, including LineageOT~\cite{forrow2021lineageot}, COSPAR~\cite{wang2022cospar} and MOSLIN~\cite{lange2023mapping}. The outputs of all three are different from ours: they use the lineage information to reconstruct couplings between observed cells, while we aim to use this information to reconstruct the path-measure of a stochastic process modeling cell differentiation, and the couplings only appear as one part of this reconstructed path-measure. Critically, like all other trajectory inference methods we are aware of, they cannot disentangle variability in proliferation rates from biases in differentiation.

%\af{Mention CoSpar? It uses a different kind of lineage tracing data, but helpfully points out that it ``cannot distinguish fate bias from differences in net rates of cell expansion (division – loss)'' (Supplement Fig. 1d)}\ev{COSPAR has been added}

\subsubsection*{Mathematical overview}

\resolved{\gs{I have attempted to reorganize this section to put the problem first, before telling what the Schrodinger problem is.}
\gs{We are given data...}
\gs{From this data we wish to recover $P^\tau$.}
\gs{Lavenant et al show that $P^\tau$ can be recovered, in the case without branching or death.}}

In this article, we model the biological process of cellular differentiation, in a time interval $[0,T]$, with a branching stochastic differential equation (branching SDE). 

More precisely, each cell is described at any time $t \in [0,T]$ by a random vector $X_t$ specifying its cell state. This vector belongs to the so-called gene expression space $\mathcal X \subset \mathbb R^g$, where $g$ is the number of genes of interest, and follows the law of an SDE. We assume that the drift of the SDE is the gradient of some potential $\Psi$ defined from $[0,T] \times \mathcal X$ to $\mathbb R$, and that the diffusion coefficient is a constant $\tau$. The spatial evolution of the cell state $X_t \subset \mathcal X$ follows
\begin{align}
\label{eq_SDE}
    \mathrm d X_t = -\nabla \Psi(t, X_t)\mathrm dt + \sqrt{\tau} \mathrm d B_t.
\end{align}
We denote by $P^\tau \in \mathcal P(\textrm{c\`adl\`ag}([0, T], \mathcal X))$ the path-measure characterizing this SDE.

Moreover, to take into account proliferation, cells can divide into two independently evolving cells or die at exponential branching rates $b(t, x)$ and $d(t, x)$ respectively. The branching SDE is then measure-valued (see Section \ref{sec_state_of_the_art} for more details), and the path-measure characterizing it belongs to $\mathcal P(\textrm{c\`adl\`ag}([0, T], \mathcal M_+(\mathcal X)))$, where $\mathcal M_+(\mathcal X)$ denotes the space of positive measures on $\mathcal X$.

The problem of trajectory inference consists in reconstructing this path-measure from a sequence of temporal marginals of the process at $N$ times $0\le t_1<\cdots<t_N\le T$, which we denote $\mu:=(\mu_{t_1},\cdots,\mu_{t_n})$. The temporal marginals are positive measures on $\mathcal X$, which do not necessary sum to one due to branching and death. In our setting, they correspond to the experimental observations performed by biologists. 

\bigskip

The methods WOT~\cite{Schiebinger2019} and gWOT~\cite{Lavenant2021} that we build on in this article both address the case where there is neither branching nor death and the path-measure is simply $P^\tau$. WOT aims to reconstruct $P^\tau$ by minimizing the \emph{relative entropy}, defined by
\begin{align}
    \label{def_entropy}
    H(R|W^\tau) := \begin{cases}
        \mathbf E^{W^\tau}\left[h\left(\frac{\mathrm dR}{\mathrm dW^{\tau}}\right)\right] &\textrm{ if }R \ll W^\tau,\\
        +\infty &\textrm{ otherwise,}
    \end{cases}
\end{align}
among all the path-measures $R\in \mathcal P(\textrm{c\`adl\`ag}([0, T], \mathcal X))$ that match the sequence of prescribed temporal marginals $\mu$. Here, $W^{\tau}$ is the path-measure of a Brownian motion with diffusivity $\tau$ and $\mathbf E^{W\tau}$ denotes the expected value under the law of $W^{\tau}$. The function $h$ is defined by\footnote{Note that the function $h$ is generally introduced without the $1-x$ because the expected value of $1 - \frac{dR}{dW^\tau}$ is always $0$ when we consider path-measures whose temporal marginals integrate to 1, which is not always the case in this article.} $h:\, x \rightarrow x\ln x + 1 - x$, and $\frac{\mathrm dR}{\mathrm dW^{\tau}}$ is the standard Radon-Nikodym derivative between the two path-measures. This minimization problem is generally known as the \emph{Schr\"odinger problem}, and has been the object of many studies in the last ten years due to its connections with optimal transport~\cite{gentil2017analogy, peyre2019computational,leonard2013survey}.

\bigskip

In practice however, the temporal marginals can only be approximated by the empirical distributions describing the sets of cells that are observed at each time of measurement. We denote $\hat \mu := (\hat \mu_{t_1},\cdots,\hat \mu_{t_N})$ the resulting sequence of empirical distributions. 

The method gWOT improves on WOT by considering the case where the number of cells characterizing each $\hat \mu_{t_i}$ is small. Lavenant et al. show that the marginal \emph{constraints} in the Schr\"odinger problem \eqref{def_entropy} should be replaced in that case by marginal \emph{penalizations}. In particular, one of their main results (Theorem 2.4 of~\cite{Lavenant2021}) is that $P^\tau$ coincides with the minimum of the functional
\begin{align}
\label{functional_FTLH}
    F_{N,\lambda,h}(\hat \mu):\, R\to\tau H(R|W^{\tau}) + \frac{1}{\lambda}\sum\limits_{i=1}^N ({t_{i+1} - {t_i}})H(\Phi_h * \hat \mu_{t_i}\big| R_{t_i}),
\end{align}
in the limit $N \to \infty$, followed by $\lambda \to 0, h \to 0$. Here, $\Phi_h$ denotes a Gaussian kernel of width $h$, $*$ stands for the convolution product, and $R_{t_i}$ are the marginals of $R$ at time $t_i$. This result can be seen as a \emph{consistency result} of the gWOT method, which consists in minimizing the functional \eqref{functional_FTLH} given a sequence of experimental observations. 

\bigskip

When we take into account cell proliferation and death, it is still possible to define the relative entropy by replacing the Brownian motion by a branching Brownian motion (BBM). The associated \emph{unbalanced Schr\"odinger problem} has been studied in depth recently~\cite{baradat2021regularized}, but there remain important obstacles to applying the new results to the theory of trajectory inference. In particular, the algorithm minimizing the functional \ref{functional_FTLH} requires solving the Schr\"odinger problem, and simple methods for finding a solution when the reference measure is a Brownian motion (like the well-known Sinkhorn algorithm~\cite{sinkhorn1967diagonal, Cuturi2013}) seem unusable in the case with branching (see~\cite{baradat2021regularized}, Section 6). Moreover, the drift associated to the resulting optimal path-measure depends explicitly on the choice of the branching rates of the reference BBM (\emph{ibid}, Section 4), which prevents us from having a consistency result similar to Theorem 2.4 of~\cite{Lavenant2021}. These limitations highlight the need for additional information to extend the theory of trajectory inference to the branching case.

\bigskip

We show in this article that the information provided by lineage tracing suffices to account for branching, at least in some particular cases. One of our main result is Theorem \ref{thm_consistency_1}, which is an extension of Lavenant et al.~\cite{Lavenant2021}, Theorem 2.4, and can be summarized as follows:

\begin{theo}
\label{thm_consistency_1_reduced}
With the previous notation, if the death rate $d = 0$, we can build, from the sequence of empirical measures $(\hat \mu_{t_i})_{i=1\cdots,N}$ and the lineage tree, a sequence of experimental probabilistic distributions $\hat p = (\hat p_{t_i})_{i=1\cdots,N}$ such that the minimizer $R_{N,\lambda,h}$ of $F_{N,\lambda,h}(\hat p)$ converges narrowly to $P^{\tau}$ in $\mathcal P(\textrm{c\`adl\`ag}([0, T], \mathcal X))$ in the limit $N \to \infty$, followed by $\lambda \to 0, h \to 0$. 
\end{theo}

\begin{figure}
	    \includegraphics[trim=0cm 0cm 0cm 9cm, width=1\textwidth]{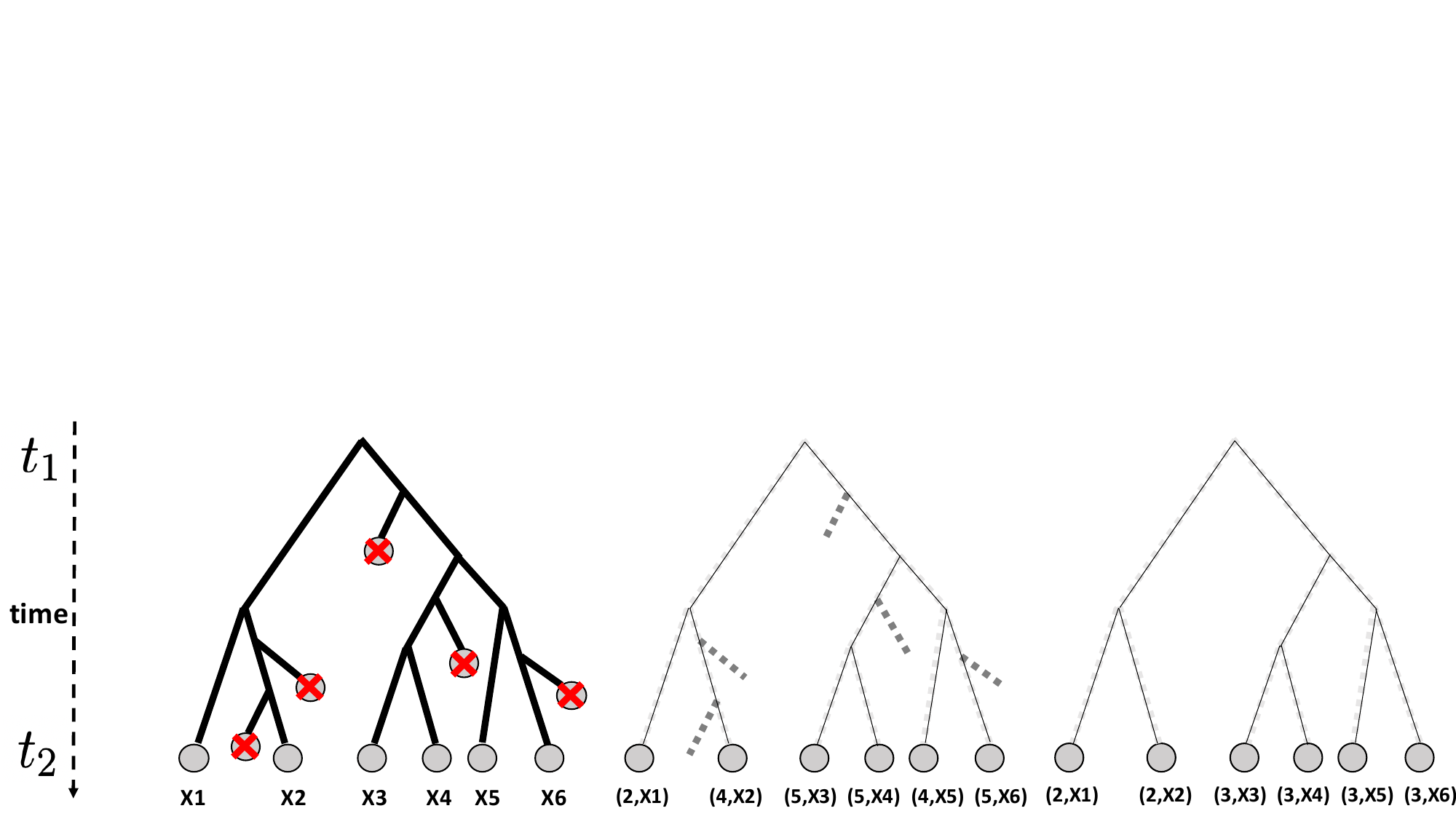}
     \textbf{Cells on}:  \hspace{.3cm} (A) $\mathcal X$ + tree structure \hspace{.7cm} (B) $\mathbb N \times \mathcal X$ using $m(X)$ \hspace{.6cm} (C) $\mathbb N \times \mathcal X$ using $\tilde m(X)$
     \caption{Different representations of a leaves of a tree evolving between $t_1$ and $t_2$, represented in (A), using: (B) the real generation numbers $m(X)$ associated to every leaf $X$ and (C) the observable generation numbers $\tilde m(X)$ associated to every leaf $X$. Note that if no cell dies between $t_1$ and $t_2$, the representations in (B) and (C) coincide.}
     \label{Figure1}
\end{figure}

A key point of interest in this result is that the convergence does not depend on the number of trees  or cells constituting the empirical observations. \resolved{\af{Why is that the main advantage? Advantage over what? The main advantage relative to gWOT is that we allow branching.}\ev{I believe the word interest is more appropriate! Interest, as detailed in the gWOT paper, that we can have guarantees without requiring an infinite number of cells at each timepoint.}} Its proof relies on a simple, important, and to the best of our knowledge never explicitly pointed out, observation: if we observe a set of trees corresponding to independent realizations of a branching process, knowing the \emph{observable generation number} $\tilde m(x)$ of each cell $x$, i.e. the number of divisions recorded in a lineage tree that cell $x$ underwent before its measurement, enables reconstructing the time-varying distribution of a new stochastic process without proliferation. This process coincides:
\begin{enumerate}
    \item \label{case_1} with the real underlying SDE \eqref{eq_SDE} when there is no death and we observe all the leaves of each observed tree;
    \item \label{case_2} with a new SDE with a modified drift when the cells can die and/or the observed cells are a subsample of the real number of cells corresponding to each tree;
    \item \label{case_3} with the real underlying branching SDE with only vanishing particles \eqref{master_equation_onlydeath} if we consider that we are able to measure real generation numbers $m(x)$ (see Fig.~\ref{Figure1}) instead of the $\tilde m(x)$ from lineage tracing.
\end{enumerate}

Case \ref{case_1} presented above is precisely the case for which Theorem \ref{thm_consistency_1_reduced} applies, after a slight extension of existing results. For case \ref{case_2}, we describe in Section \ref{sec_method_bias} the bias arising in this situation, which depends on the survival probabilities of the branching process with death and subsampling. Our main finding is that provided that every initial cell at time $0$ has at least one observed descendant at every timepoint, the reweighted empirical measures \eqref{reweighted_empirical_measures} converges, in the limit of an infinite number of trees, to the intensity of an SDE with a biased drift that follows the system of master equation \eqref{SDE_withbias} (see Corollary \ref{cor_conditionalm}). We also present a numerical method for reducing the bias using the times of last common ancestors for each pair of leaves of the lineage trees.

\bigskip

Finally, we state a second consistency theorem handling case \ref{case_3}, showing that observation of the real generation numbers could let us remove the bias due to subsampling, at least in the case when the death rate is uniformly $0$.\resolved{\nitya{Doesn't making the death rate zero just simplify this to Case 1?}\ev{Only if there's no subsampling, if not it's stronger}} When $d>0$, accurate prior knowledge of the death rate is still required for the consistency (see Theorem \ref{thm_consistency_2}). Although experimental data with real generation numbers are not currently accessible in the literature, they are in line with current development of measurement technologies~\cite{masuyama2022molecular}.

%\gs{We could also elevate Corollary 11, which describes the bias if there are inexact generation numbers, to a main result.}
%\ev{I decided No at the moment, and to only provide an explicit reference to the corollary, but it could be discussed!}
%\af{I agree about not elevating it. I'd just highlight that the bias is (1) mathematically described and (2) in practice (and in theory?) not very big.}

\subsection*{Organization of the paper}
In Section \ref{sec_state_of_the_art}, we describe the experiments providing single-cell data with lineage tracing, and detail the subsequent mathematical assumptions. In Section \ref{sec_deconvolution}, we present the main idea of this article, that is to use generation numbers to \emph{deconvolve} the proliferation, \emph{i.e.} to find the distribution of the underlying stochastic process with only diffusing and vanishing cells. In Section \ref{sec_consistencytheorem1} we present the theoretical guarantees associated to the case of null death rate and no subsampling. We illustrate how our deconvolution method compares with the heuristic method for trajectory inference with branching presented in Chizat et al.~\cite{zhang2022trajectory}, which required the knowledge of the branching rates instead of lineage tracing, and show that our results also allow for an efficient numerical reconstruction of the drift and the birth rate characterizing the branching SDE. Section \ref{sec_method_bias} is devoted to the analysis of the bias which appears in case of death and subsampling, and how to reduce it. Finally, we prove in Section \ref{sec_death_process} that the observation of the real number of divisions could permit removing this bias, using accurate prior knowledge of the death rate alone.

\section{Experimental setting and mathematical assumptions}
\label{sec_state_of_the_art}

It has long been impossible in cell biology to generate experimental datasets such that both the gene expression at the single-cell level and information about cellular relationships, as the lineage tree between the measured cells, are available. But these last few years, measurement technologies have seen tremendous recent advances, and it is now possible to recover the full lineage tree of a population of cells~\cite{mckenna2016whole, raj2018simultaneous}. The aim of this section is to briefly describe the experimental setting where these data are obtained and to lay out our subsequent mathematical hypotheses.

\subsection{Description of the experimental setting}

Technologies for reconstructing cellular lineage trees use CRISPR–Cas9 genome editing technology to continuously mutate an array of synthetic DNA barcodes. These barcodes are incorporated into the chromosomes so that they are inherited by daughter cells. They are then further mutated over the course of development, in such a way that when a population of cells are measured with RNA-sequencing, analyzing the pattern of mutations in the barcodes of each cell allows reconstructing a lineage tree which describes shared ancestry within the population. Moreover, as DNA barcodes are expressed as transcripts, they can be simply recovered together with the rest of the transcriptome with scRNA-seq. 

As is usually the case with RNA-sequencing, cells must be lysed before information about their state or lineage is recovered. Thus, this measurement technology is destructive: the data at each timepoint are independent in that a cell observed at a given timepoint can only share ancestors with cells observed at the same timepoint.

The data is therefore a sequence of independent arrays (one for each timepoint). At each timepoint $t_i$, the corresponding array is of size $N_i \times g$ where $N_i$ is the number of cells observed, perhaps of order $1e3-1e4$, and $g$ the number of genes observed, typically $\sim 1e4$. Each coordinate of this array corresponds to the number of ``reads" of mRNAs that are measured. 

Moreover, the barcodes attached to the reads which enable identification of the cells in classical RNA-sequencing technology here also enable reconstruction of the lineage tree of shared ancestry. The problem of reconstructing the lineage trees from these mutated barcodes is itself a challenge, for which recent tools have shown high efficiency~\cite{Chan2019,weinreb2020lineage}, including when the number of cells is very high~\cite{konno2022deep}. We assume in this paper that we have access at each timepoint $t_i$ to a lineage tree associated to the cells that are measured, that we denote $\mathcal T_{t_i}$ in the following, and we focus on using this information for trajectory inference. 

\bigskip

Importantly, although at each measured timepoint we have a tree associated to a population of cells, the information provided about both this tree and the positions of its leaves in the gene expression space is not complete. First, only a fraction of the reads expressed by a cell are sampled, which induces noise in the observation of the cellular states. Second, divisions may happen more often than barcodes mutate, generating some errors in the tree reconstruction. Third, because only a small fraction of the descendants of the cells initially present in the experiment are sampled at each timepoint, we observe only a subset of the nodes in the true lineage tree. The two first problems are likely to generate some noise in the data, the analysis of which is out of the scope of this article. The first one in particular is a classical problem in scRNA-seq data for which standards preprocessing methods have been developed~\cite{seurat2018}, and we expect the second one to be considered in the same way. The third problem, called \emph{subsampling} in the following, is specific to lineage tracing technologies and, to the best of our knowledge, its implications have never been properly studied. We will show that it induces a bias in the trajectory inference that we will study carefully in Section \ref{sec_method_bias}.

\subsection{Description of the mathematical setting}

We consider that the time-course of gene expression profiles, which we observe together with their lineage trees at an increasing sequence of timepoints $\{t_1, \cdots, t_N\}$, corresponds to independent realizations of a branching SDE with values on the gene expression space $\mathcal X \subseteq \mathbb R^g$, on which we have subsampled a certain number of cells. We denote the last time of measurement by $T:=t_N$.

Ignoring for the moment the lineage tree and considering only the leaves of each tree, the data can be described by the superposition of two processes: first, a measure-valued process of law $P \in \mathcal P(\textrm{c\`adl\`ag}([0, T], \mathcal M_+(\mathcal X)))$, corresponding to the branching SDE; and second, a subsampling process on the realizations of the branching SDE at the observed timepoints. We will describe this model first, including how the subsampling is taken into account. In the next section we will detail how the lineage tree is taken into account through the generation numbers.

\subsubsection*{Description of the model}

The branching SDE process we consider has two main characteristics:
\begin{itemize}
    \item \textbf{The spatial motion}: During their lifetime, each cell moves around in $\mathcal X$, independently of the other cells, following a SDE of the form \eqref{eq_SDE};
    \item \textbf{The branching mechanism}: Each cell has an lifetime which is independent from the other cells, and is exponentially distributed: given that a cell is alive in $x$ at time $t$, it divides into two cells at time $t + \delta t$ with probability $b(t, x)\delta t + o(\delta t)$, and dies with probability $d(t, x)\delta t + o(\delta t)$.
    When it divides at $x$, it gives rise to two cells at $x$ that evolve independently.
\end{itemize}
As a consequence, a branching SDE satisfies these two properties, that are key for its analysis:
\begin{itemize}
    \item \textbf{The Markov property}: Let $(\mathcal F_t)_{t \in [0,T]}$ be the natural filtration associated to the process, let $t \in [0,T]$, and let $t^*$ be a stopping time with respect to this filtration such that $t^* \leq t$ $P$-a.s. For any measurable function $F$ from $\mathcal M_+(\mathcal X)$ to $\mathbb R^+ \cup \{+\infty\}$, we have:  \begin{align}
    \label{markov_property}
    \mathbf E^P\left[F(Z_t)\big| \mathcal F_{t^*}\right] = \mathbf E^P_{Z_{t^*}}\left[F(Z_{t - t^*})\right],
\end{align}
where $Z_s$ is a random variable describing the sum of Dirac masses on $\mathcal X$ characterizing the process at a time $s$, which then follows the law of $P$. $\mathbf E^P_{Z_{t^*}}\left[\cdot\right]$ denotes the expected value, under the law of $P$, starting from the initial condition $Z_0 = Z_{t^*}$.
\resolved{\af{I'm not totally following this formulation of the Markov property. Do we really want $F$ to be a function of the measure $Z_t$?}\ev{Should be solved now}}
    \item \textbf{The branching property}: Let $t \in [0,T]$, and denote $\mathcal N(t) = \textrm{Supp}\,Z_t$. For any sum of Dirac masses $\mu$ in $\mathcal X$ and any measurable function $u$ from $\mathcal X$ to $\mathbb R^+ \cup \{+\infty\}$, we have:
    \begin{align}
    \label{branching_property}
    \mathbf E^P\left[\prod\limits_{X \in \mathcal N(t)}u(X)\big| Z_0 = \mu\right] = \prod\limits_{x \in \textrm{Supp}\,\mu} \mathbf E^P\left[\prod\limits_{X \in \mathcal N(t)}u(X)\big| Z_0 = \delta_x\right].
\end{align}
\end{itemize}

\bigskip

We can define a notion of \emph{intensity} for such processes as follows:
\begin{defn}
\label{first_def_intensity}
We define the intensity of the branching SDE, $\mathbf E^P[Z_t]$, as the measure defined on the Borel sets $A \subset \mathcal X$ as: $\mathbf E^P[Z_t]:\, A \to \mathbf E^P[Z_t(A)]$.
\end{defn}
Equivalently, under assumption \eqref{assumption_rates}, $\mathbf E^P[Z_t]$ is the measure of $\mathcal M_+(\mathcal X)$ defined for any function $\theta \in C^b(\mathcal X)$ by
\begin{align*}
    \langle \theta, \mathbf E^P [Z_t] \rangle = \mathbf E^P\left[ \langle\theta, Z_t\rangle\right],
\end{align*}
where $\langle \cdot, \cdot \rangle$ stands for the duality bracket between continuous functions and measures of finite total variation in $\mathcal X$.

\bigskip

Throughout our article, we will work under the following very classical assumption that the branching mechanisms $b, d$  are uniformly bounded,
\begin{conj}
\label{assumption_rates} $||b + d||_{\infty} < \infty$,
\end{conj}
which ensures that the number of cells at each timepoint is almost surely finite. We don't mention this assumption in the following, since it is implicit in all our results.

As we are going to prove relations between the time-varying intensity of a branching SDE and the probabilistic distribution of its underlying SDE, it will also be simpler to consider that the initial measure characterizing the process is probabilistic, i.e. has total mass 1:
\begin{conj}
\label{assumption_initial_distrib}
The random variable $Z_0$ is sampled under a probability law: $$\mathbf E^P[Z_0] = \mu_0 \in \mathcal P(\mathcal X).$$
\end{conj}
This is in line with the process we observe since life starts with the formation of a single egg. However, we emphasize that our results can be easily extended to the case where $\mu_0(\mathcal X) \geq 1$ by multiplying temporal marginals of the reconstructed path-measures by $\mu_0(\mathcal X)$. 

\bigskip

We denote in the following $\mathbf E^P_x[\cdot]$ the expected value under the law of $P$ conditionally to $Z_0 = \delta_x$, and $\mathbf E^P[\cdot]$ when the initial condition is the probabilistic distribution $\mu_0$ given by Assumption \ref{assumption_initial_distrib}. When there is no confusion on the law on which we consider the expected values, we omit the $P$ in the expected value.

\subsubsection*{Effect of subsampling}

We consider that the subsampling at a time $t \in [0,T]$ consists in taking each cell in $\mathcal N(t)$ with a probability $q(t)$, constant in $\mathcal X$. Importantly, we assume that this probability does not depend on the position of the cells in $\mathcal X$, nor on the number of cells in $\mathcal N(t)$. We do allow this probability to depend on time, to take into account the fact that we expect a higher proportion of the cells to be subsampled at early times, when the number of cells is low, than later when the true number of cells is expected to be very high. Thus, the subsampling is just an additional layer: denoting by $\tilde P$ the modified path-measure characterizing the process described by $P$ with subsampling, for any Borel sets $A \subset \mathcal X$ we have:
\begin{align}
\label{rel_intensity_subsampling}
 \mathbf{E}^{\tilde P}[Z_t](A) = q(t) \mathbf E^{P}[Z_t](A).
\end{align}

\subsubsection*{Inverse problem with lineage tracing}

With this model in hands, we can rigorously state the problem of trajectory inference from scRNA-seq data with lineage tracing. It is the question of reconstructing the path-measure $P$ from a sequence of experimental measures $\hat \mu:=\{\hat \mu_{t_i}\}_{i=1\cdots,N}$ that are assumed to be sampled under the temporal marginals of $\tilde P$, together with a sequence of lineage trees $\{\mathcal T_{t_i}\}_{i=1,\cdots,N}$ describing at each timepoint the shared ancestry of the cells constituting the associated empirical measure.

The first and main challenge arising from this inverse problem is how to use the lineage tree information to formulate a minimization problem that characterizes the path measure $P$ in the limit $N \to \infty$. As mentioned in the introduction, we show in this article that the generation numbers from the lineage tree suffice for stating and solving such a minimization problem, at least under some conditions. We are now going to detail how to take into account these numbers in a dynamical way,\resolved{\af{I'm not sure what you mean by ``probabilistic'' here.}\ev{I said probabilistic instead of dynamical, to emphasize that we are going to take it into account in the stochastic process directly}} by integrating them to the description of the branching process.

\section{Deconvolving the proliferation using generation numbers}
\label{sec_deconvolution}

In this section, we extend the description of a branching SDE presented in Section~\ref{sec_state_of_the_art} to the case where generation numbers are observed. 
To make the mathematical presentation cleaner, we start by considering the real generation numbers (Fig.~\ref{Figure1}.B), and postpone discussion of the partial generation numbers coming from lineage tracing (Fig.~\ref{Figure1}.C) to Section~\ref{sec_method_bias}. The two types of generation numbers coincide only for experiments without subsampling or death.

The main result of this section could be generalized to any branching process with only birth and death mechanisms, not only branching SDEs.

\subsection{Mathematical description and master equation of a branching SDE with real generation numbers}

For the branching SDE with real generation numbers, the SDE \eqref{eq_SDE} still characterizes the spatial motion, and the rates $b, d$ the branching mechanism. The only difference is that a counter is attached to each leaf of the tree recording the number of divisions the cell has been through. For any time $t \in [0,T]$, $\mathcal N(t)$ now denotes a set of cells each described by a tuple $(m, x)$ containing the generation number $m\in\mathbb N$ and position $x\in\mathcal X$. The process can thus be described by a family of random variables $Z_t:= \{Z_t(m)\}_{m\in \mathbb N}$ each describing the empirical measure on $\mathcal X$ of the process for a given generation number $m$:
$$Z_t(m) = \sum\limits_{(m, x) \in \mathcal N(t)}\delta_{x}.$$
A realization of a branching SDE is then a $\textrm{c\`adl\`ag}$ curve valued in $\mathcal M_+(\mathcal X)^{\mathbb N}$, the jumps of which correspond to the branching events. 

\bigskip

The intensity of the branching SDE with real generation numbers is defined as a natural extension of \eqref{first_def_intensity}. Let $P$ be the path-measure of the branching process with real generation numbers. For any $m \in \mathbb N$, the intensity at time $t$, written $\mathbf E^P[Z_t(m)]$, is the measure on $\mathcal X$ defined by
\begin{align}
    \label{def_intensity_counts}
    \langle \theta, \mathbf E^P [Z_t(m)] \rangle = \mathbf E^P\left[ \langle\theta, Z_t(m)\rangle\right]
\end{align}
for any test function $\theta \in C(\mathcal X)$.

The initial condition of Assumption \ref{assumption_initial_distrib} then becomes:
\begin{align*}
    \begin{cases}
        \mathbf E^P[Z_0(m)] = \mu_0 &\textrm{if }m = 0,\\
        \mathbf E^P[Z_0(m)] = 0 &\textrm{otherwise}.
    \end{cases}
\end{align*}

Our next goal is to derive a system of master equations characterizing the family of intensities $(\mathbf E^P[Z_t(m)])_{m \in \mathbb N}$.
We start from the classical result that if the potential $\Psi$ has a gradient $\nabla \Psi$ that is globally Lipschitz in $\mathcal X$, the probability distribution characterizing the SDE \eqref{eq_SDE} solves the following partial differential equation (PDE) in the weak sense:
\begin{align}
\label{master_equation_SDE}
    \partial_t p = &div(p \nabla \Psi) + \frac{\tau}{2}\Delta p := \mathcal L^*_{\Psi, \tau} p.
\end{align}
Moreover, the intensity of the branching SDE described in Section \ref{sec_state_of_the_art} solves the following PDE in the weak sense:
\begin{align}
\label{master_equation_BBM}
    \partial_t \rho = \mathcal L^*_{\Psi, \tau} \rho + (b - d)\rho.
\end{align}
A proof of this claim can be found for example in~\cite{baradat2021regularized} (Corollary 2.39). \resolved{\af{The reference does have a proof, right?}\ev{It presents the main ideas of the proof in the main text, similar to what I do in this article for the process with m, and a completely rigorous proof can be derived from the appendix (which they refer to in the main text).}}
\bigskip

For the branching SDE with real generation numbers, we prove the following proposition:
\begin{prop}
\label{prop_master_equation_realm}
The family of intensities $\{\mathbf E^P[Z_t(m)]\}_{m \in \mathbb N}$ solves the following system of PDEs in the weak sense:
\begin{align}
\label{master_equation_system}
    \forall m \geq 0:\, \partial_t \rho(m) = \mathcal L^*_{\Psi, \tau}\rho(m) + 2b \rho(m-1) - (b + d)\rho(m),
\end{align}
with the convention $\rho(-1, \cdot) = 0$, and initial condition $\rho_0(m) = \mu_0\mathds 1_{m=0}$.
\end{prop}

Before detailing the proof, we remark that as expected, the sum $\sum\limits_{m \geq 0} \rho(m)$ solves Eq.~\eqref{master_equation_BBM}. Moreover, this equation can be understood as follows:
\begin{itemize}
    \item At any time, the spatial motion characterizing the branching SDE with real generation number is the same as for the branching SDE;
    \item When the exponential clock of a cell in $(m, x)$ rings, it necessarily dies or gives rise to two cells in $(m+1, x)$.
\end{itemize}

We only give the main elements of the proof, since the details are similar to the ones that can be found in other more complete references, including~\cite{baradat2021regularized},~\cite{li2011measure} and~\cite{etheridge2000introduction}. We nevertheless give all the main steps that will allow us to extend the result to conditional processes in the next section. The proof relies mainly on the Markov and the branching properties (\eqref{markov_property} and \eqref{branching_property}). 
%It also crucially uses the stability of the process by translation in $\mathbb N$, that we can express as follows. Let $(m,x) \in \mathbb N \times \mathcal X$ , $t \in [0, T]$, and $\{\theta(m)\}$ be a family of measurable functions from $\mathcal X$ to $\mathbb R$: we have for all $m' \geq 0$:
%\begin{align}
%\label{invariance_truem}
    %\sum\limits_{m' \geq 0}\mathbf E_{{(m,x)}}\left[\langle \theta(m'), Z_t(m') \rangle \right] = \sum\limits_{m' \geq 0} \mathbf E_{{(0,x)}}\left[\langle \theta(m' + m), Z_t(m') \rangle \right],
%\end{align}

\begin{proof}[Proof of Proposition~\ref{prop_master_equation_realm}]
The aim is to prove that for any family of test functions in $\mathcal X$, $\{\theta(m)\}_{m\in \mathbb N}$, the family of intensities $\{\mathbf E^P[Z_t(m)]\}_{m \in \mathbb N}$ solves the following PDE:
\begin{align}
\label{master_equation_dual}
    \sum\limits_{m \in \mathbb N}\frac{\mathrm d}{\mathrm dt}\langle \theta(m), \rho(m) \rangle =  \sum\limits_{m \in \mathbb N} \bigg(&\langle \mathcal L_{\Psi, \tau} \theta(m), \rho(m) \rangle \\&+  \langle\theta(m+1), 2b\rho(m) \rangle -  \langle \theta(m), (b+d)\rho(m) \rangle\bigg),
    \notag
\end{align}
where $\mathcal L_{\Psi, \tau}$ is the generator of the underlying SDE \eqref{eq_SDE}: 
$$\mathcal L_{\Psi, \tau} \theta := -\langle \nabla \Psi, \nabla \theta\rangle + \frac{\tau}{2} \Delta \theta.$$
It suffices to consider $\theta(m) > 0$, as any $\theta(m)$ could be decomposed into $\theta(m)^+ - \theta(m)^-$ with $\theta(m)^+>0$ and $\theta(m)^->0$.

Let $u_0$ be a function from $\mathbb N \times \mathcal X$ to $[0, 1]$ such that for all $m \in \mathbb N$, $u(m, \cdot)$ is smooth. Let $x \in \mathcal X$ and $m \in \mathbb N$. Thanks to the branching property, we can restrict to the case where $\mu_0 = \delta_{(m, x)}$. The proof consists of three steps: (1) find the derivative in $t = 0$ of $u(t, m, x) := \mathbf E_{(m,x)}\left[\prod\limits_{(M, X) \in \mathcal N(t)} u_0(M, X)\right]$, (2) deduce that $u$ is the classical solution of a certain PDE for any $u_0$, and (3) deduce that the family of intensities defined by \eqref{def_intensity_counts} is a weak solution of the PDE \eqref{master_equation_dual}.

\bigskip

\underline{Step 1}: Starting from a cell in $x$, at a small time $\delta t$, the probability that the cell has divided is $b \delta t + o(\delta t)$, the probability that that it is dead is $d \delta t + o(\delta t)$, and the probability that no branching event happened is $1 - (b+d) \delta t + o(\delta t)$. Any other event has a probability in $o(\delta t)$. Thus, denoting $A^b, A^d$ and $A^c$ these three complementary events, we have:
\begin{align*}
u(\delta t, m, x) &= (1 - (b+d)\delta t) \mathbf E_{(m, x)}\left[\prod\limits_{(M, X) \in \mathcal N(\delta t)} u_0(M,X) \big| A^c\right] \\ &+ b \delta t \mathbf E_{(m, x)}\left[\prod\limits_{(M, X) \in \mathcal N(\delta t)} u_0(M,X) \big| A^b\right] + d \delta_t \mathbf E_{(m, x)}\left[\prod\limits_{(M, X) \in \mathcal N(\delta t)} u_0(M,X) \big| A^d\right] + o(\delta t),\\
 &= (1 - (b+d)\delta t) \mathbf E_x^p\left[u_0(m, X_{\delta t})\right] + b \delta t \left(\mathbf E_x^p\left[u_0(m+1, X_{\delta t})\right]\right)^2 + d \delta_t + o(\delta t),
\end{align*}
where $\mathbf E_x^p$ denotes the expected value under the law of the underlying SDE \eqref{eq_SDE}, which is a probabilistic process. The passage from the first to the second line is justified by the branching property. We then obtain:
\begin{align*}
\lim\limits_{\delta t \to 0} \frac {u(\delta t, m, x) - u_0(m, x)}{\delta t} = \lim\limits_{\delta t \to 0} \frac {\mathbf E_x^p\left[u_0(m, X_{\delta t})\right] - u_0(m, x)}{\delta t} - (b+d) u_0(m, x) + b u_0(m+1, x)^2 + d.
\end{align*}
The limit on the right-hand side being equal to the generator $\mathcal L_{\Psi, \tau}$ applied to $u_0(m, x)$, we then obtain the following system of equations:
\begin{align*}
\forall m \in \mathbb N, \forall x \in \mathcal X:\, \frac{\mathrm d}{\mathrm d t}u(t, m, x)\big|_{t = 0} &= \mathcal L_{\Psi, \tau}  u_0(m,x) + b u_0(m+1, x)^2 + d - (b+d) u_0(m, x),\\ &:= \mathcal K_{\Psi, \tau} u_0.
\end{align*}

\underline{Step 2}: Using the Markov and the branching properties of the branching process, we have for all $t, s > 0$:
\begin{align*}
u(t+s, m, x) &= \mathbf E_{(m, x)}\left[\mathbf E_{(m, x)}\left[\prod\limits_{(M, X) \in \mathcal N(t+s)} u_0(M,X) \big| \mathcal F_s\right] \right], \\ &= \mathbf E_{(m, x)}\left[\mathbf E_{Z_s}\left[\prod\limits_{(M, X) \in \mathcal N(t)} u_0(M,X)\right] \right], \\ &= \mathbf E_{(m, x)}\left[\prod\limits_{(M, X) \in \mathcal N(s)} u(t, M, X)\right].
\end{align*}
Thus, thanks to Step 1, we obtain the following system of PDEs:
\begin{align}
\label{equation_K_inter}
\forall m \in \mathbb N, \forall x \in \mathcal X:\, \frac{\mathrm d}{\mathrm d s}u(t+s, m, x)\big|_{s = 0} = \mathcal K_{\Psi, \tau} u(t, m, x).
\end{align}

\underline{Step 3}: 

Taking $\varepsilon \ll 1$ such that we can define, for all $m \in \mathbb N$, $u_0(m) := 1 − \varepsilon \theta(m) \in C(\mathcal X, [0, T])$, we obtain the following relations:
\begin{align*}
    &u(t, m, x) = 1 - \varepsilon  \mathbf E_{(m,x)}\left[\sum_{m'} \langle \theta(m'), Z_t(m')\rangle \right]+ o (\varepsilon),\\
    &\mathcal L_{\Psi, \tau} u(t, m, x) = −\varepsilon \mathcal L_{\Psi, \tau} \mathbf E_{(m,x)}\left[\sum_{m'} \langle \theta(m'), Z_t(m')\rangle \right]+ o (\varepsilon),\\
    &b u(t, m+1, x)^2 + d - (b+d) u(t, m, x)  = b + d - (b + d) \\&\hspace{2cm}-\varepsilon\left(2b\mathbf E_{(m+1,x)}\left[\sum_{m'} \langle  \theta(m'), Z_t(m')\rangle \right] - (b+d) \mathbf E_{(m,x)}\left[\sum_{m'} \langle\theta(m'), Z_t(m')\rangle\right]\right) + o(\varepsilon).
\end{align*}

Thus, we obtain by substituting in \eqref{equation_K_inter} the terms appearing on the left-hand side of the three previous relations by the right-hand side formulas:

\begin{align*}
\frac{\mathrm d}{\mathrm dt} \mathbf E_{(m,x)}\left[\sum_{m'} \langle \theta(m'), Z_t(m')\rangle \right] =  &\mathcal L_{\Psi, \tau} \mathbf E_{(m,x)}\left[\sum_{m'} \langle \theta(m'), Z_t(m')\rangle \right]  +2b\mathbf E_{(m+1,x)}\left[\sum_{m'} \langle  \theta(m'), Z_t(m')\rangle \right] \\&-(b+d) \mathbf E_{(m,x)}\left[\sum_{m'} \langle\theta(m'), Z_t(m')\rangle\right],\\
:= &\tilde{\mathcal K}_{\Psi, \tau}\mathbf E_{(m,x)}\left[\sum\limits_{m'}\langle \theta(m'), Z_t(m')\rangle \right].
\end{align*}

By the Hille-Yosida theorem, we have then:
$$\tilde{\mathcal K}_{\Psi, \tau} \mathbf E_{(m,x)}\left[\sum\limits_{m'}\langle \theta(m'), Z_t(m')\rangle \right] = \mathbf E_{(m,x)}\left[\sum\limits_{m'}\langle \tilde{\mathcal K}_{\Psi, \tau} \theta(m'), Z_t(m')\rangle \right],$$

Thus, using Definition \eqref{def_intensity_counts}, we can finally conclude that for all initial condition $\delta_{(m, x)}$, the family of intensities defined by \eqref{def_intensity_counts} solves the PDE \eqref{master_equation_dual}.
\end{proof}

\subsection{Reconstruction of the temporal marginals of a branching SDE without proliferation}

The following corollary of Proposition \ref{prop_master_equation_realm}, although very simple, is at the core of our work:

\begin{cor}
\label{cor_realm}
Let the family $\{\rho(m)\}_{m\in \mathbb N}$ be a weak solution of the system \eqref{master_equation_system}, and let us denote $\bar \rho = \sum\limits_{m \geq 0} \frac{1}{2^m} \rho(m)$. Then $\bar \rho$ solves in the weak sense the following PDE:
\begin{align}
\partial_t \bar\rho = \mathcal L^*_{\Psi, \tau} \bar\rho - d\bar \rho.
\label{master_equation_onlydeath}
\end{align}
\end{cor}

\begin{proof}
    The proof of this corollary is a straightforward application of equation \eqref{master_equation_system}. Indeed, we observe that for all $M > 0$:
    $$\partial_t \sum\limits_{m =0}^M \frac{1}{2^m} \rho(m) = \mathcal L^*_{\Psi, \tau} \left(\sum\limits_{m =0}^M \frac{1}{2^m} \rho(m)\right) - d\sum\limits_{m =0}^M \frac{1}{2^m} \rho(m) - \frac{b\rho(M)}{2^M}.$$
    Under Assumption \ref{assumption_rates}, for any $t < T$ and all $m$ the mass of the intensity $\rho(m,t, \mathrm dx)$ is finite, and $\frac{b}{2^m}\rho(m,t, \mathrm dx)$ then converges weakly to $0$ as $m \to \infty$.
\end{proof}

In plain words, the observation of generation numbers allows us to deconvolve the proliferation of cells. In particular, under Assumption \ref{assumption_initial_distrib}, if the death rate is null, the time-varying distribution $\sum\limits_{m =0}^M \frac{1}{2^m} \rho(m)$, which we call the \emph{reweighted distribution} in the remainder of the paper, coincides with the distribution of the underlying SDE \eqref{eq_SDE}. This result allows us to state the first theorem of trajectory inference for lineage tracing data. 
 
\section{Trajectory inference from lineage tracing data with neither death nor subsampling}
\label{sec_consistencytheorem1}

In this section, we present our first analog of the main theorem developed in Lavenant et al.~\cite{Lavenant2021}. Theorem~\ref{thm_consistency_1} below provides guarantees for trajectory inference using single-cell data with lineage tracing in the case where the death rate is uniformly zero and there is no subsampling. Following the proof of the theorem in Section~\ref{subsec_consistencytheorem1}, we describe how to adapt the mean field Langevin approach of~\cite{chizat2022mean} into a computationally tractable algorithm for recovering trajectories as the drift and branching rates of the underlying process.

\subsection{Consistency theorem}
\label{subsec_consistencytheorem1}

% Before to present the main result of this section, remark that by a law of large numbers on branching process, if we observe at a given time a number $n$ of cells together with their generation numbers coming from $K$ different trees at a time $t$, provided that we observe all the leaves alive at time $t$ of each tree, the empirical distribution
%\begin{align}\label{empirical_intensity}
%\hat \rho(m) = \frac{1}{K} \sum_{c = 1}^n \delta_{(m, x_c)}
%\end{align}
%converges weakly to the intensity $\mathbf E^P[Z_t(m)]$  as $K\to\infty$. 
%\af{This remark feels out of place. Do we consider the limit $K\to\infty$ anywhere else? If not, perhaps remove the remark. If it is important, it could be put where it's needed or with the definition of $\mathbf E^P[Z_t(m)]$}
%\bigskip

We consider that $K_i$ trees at each timepoint $t_i$ are sampled independently from the branching process with $d=0$, and for each tree we observe every leaf.
As presented in the introduction, our first and main theorem states that in such circumstances, we can reconstruct the path-measure of the underlying SDE \eqref{eq_SDE} from lineage tracing when the sequence of timepoints tends to be dense in $[0,T]$.
When no leaves are unobserved, the real generation numbers and the observable generation numbers $m$ and $\tilde m$ are the same (see Fig.~\ref{Figure1}); to keep expressions cleaner, we keep the notation $m$.

\begin{theo}
\label{thm_consistency_1}
Let $P^{\tau}$ be the path-measure associated to the SDE \eqref{eq_SDE}. and $W^{\tau}$ be the law of a Brownian motion with diffusivity $\tau$.  For all timepoints $t_i$, let
\begin{align}
\label{reweighted_empirical_measures}
    \hat p_{t_i} = \frac{1}{K_i}\sum\limits_{k=1}^{K_i}  \left(\sum\limits_{j=1}^{n_{k, i}}\frac{1}{2^{ m_{j,k, i}}}\delta_{x_{j,k,i}}\right),
\end{align}
where $K_i > 0$ is the number of trees observed at $t_i$ and the family of tuples $\{m_{j,k, i}, x_{j,k,i}\}_{j=1,\cdots,n_{k,i}}$ contain the generation number and gene expression for each leaf $j$ from the $k$-th tree observed at time $t_i$, all generated by the branching SDE with diffusivity $\tau$, gradient drift $v = -\nabla \Psi$, birth rate $b$ and uniformly zero death rate. 
Let $R_{N,\lambda,h}$ denote the minimizer of $F_{N,\lambda,h}(\hat p)$ defined by \eqref{functional_FTLH}.

Then, in the limit $N \to \infty$, followed by $\lambda \to 0, h \to 0$, $R_{N,\lambda,h}$ converges narrowly to $P^{\tau}$ in $\mathcal P(\textrm{c\`adl\`ag}([0, T], \mathcal X))$.
\end{theo}

\begin{proof}
This proposition follows from combining the work of Lavenant et al.~\cite{Lavenant2021} with Corollary \ref{cor_realm}. By the law of large numbers, the following weak convergence holds: for any continuous function $a$ from  $\mathbb N \times \mathcal X$ to $\mathbb R$ and every collection of non-negative weights $\{\omega_{i}^N\}_{i=1,\cdots,n}$, we have
$$\lim_{N \to \infty} \sum\limits_{i = 1}^N \omega_{i}^N \sum\limits_{m=0}^\infty\int_{\mathcal X}a(m, x)\hat \mu_{t_i}^h(m, \mathrm dx) = \sum\limits_{m=0}^\infty \int_{\mathcal X}a(m,x)\Phi_h * \rho_t(m, \mathrm dx),$$
where $\hat \mu^h_{t_i}(m, \mathrm dx) := \Phi_h * \frac{1}{K_i} \left(\sum\limits_{k=1}^{K_i} \sum\limits_{j=1}^{N_{k, i}} \delta_{(m_{j,k,i}, x_{j,k,i})}\right)$ denotes the empirical observations, and $\rho$ is the weak solution of the PDE system \eqref{master_equation_system} with initial condition $\mu_0 \mathds 1_{m=0}$.

Thus, for any continuous function $f$ from $\mathcal{X}$ to $\mathbb{R}$, we may choose $a(m,x) = \frac{f(x)}{2^m}$ to obtain the following weak convergence:
\begin{align}
\lim_{N \to \infty} \sum\limits_i^{N} \omega_i^N \int_{\mathcal X}f(x)\sum\limits_{m=0}^\infty\frac{1}{2^m}\hat \mu_{t_i}^h(m, \mathrm dx) =  \int_{\mathcal X}f(x)\sum\limits_{m=0}^\infty\frac{1}{2^m}\Phi_h*\rho_t(m, \mathrm dx).
\label{weak_convergence_TCL}
\end{align}
We next note that $\sum\limits_{m=0}^\infty\frac{1}{2^m}\hat \mu_{t_i}^h = \Phi_h * \hat p_{t_i}$. Thanks to Corollary \ref{cor_realm}, under Assumption \ref{assumption_initial_distrib}, $\sum\limits_{m=0}^\infty\frac{1}{2^m}\Phi_h * \rho_t(m, \mathrm dx) = \Phi_h * p_t(\mathrm dx)$, where $p$ is the weak solution of the PDE \eqref{master_equation_SDE} with initial condition $\mu_0$. Theorems 4.4, 4.6 and 4.1 of~\cite{Lavenant2021} then apply, and allow us to conclude.
\end{proof}

\begin{figure}
\centering
\includegraphics[trim=3cm 0cm 3.5cm 0cm, width=\textwidth]{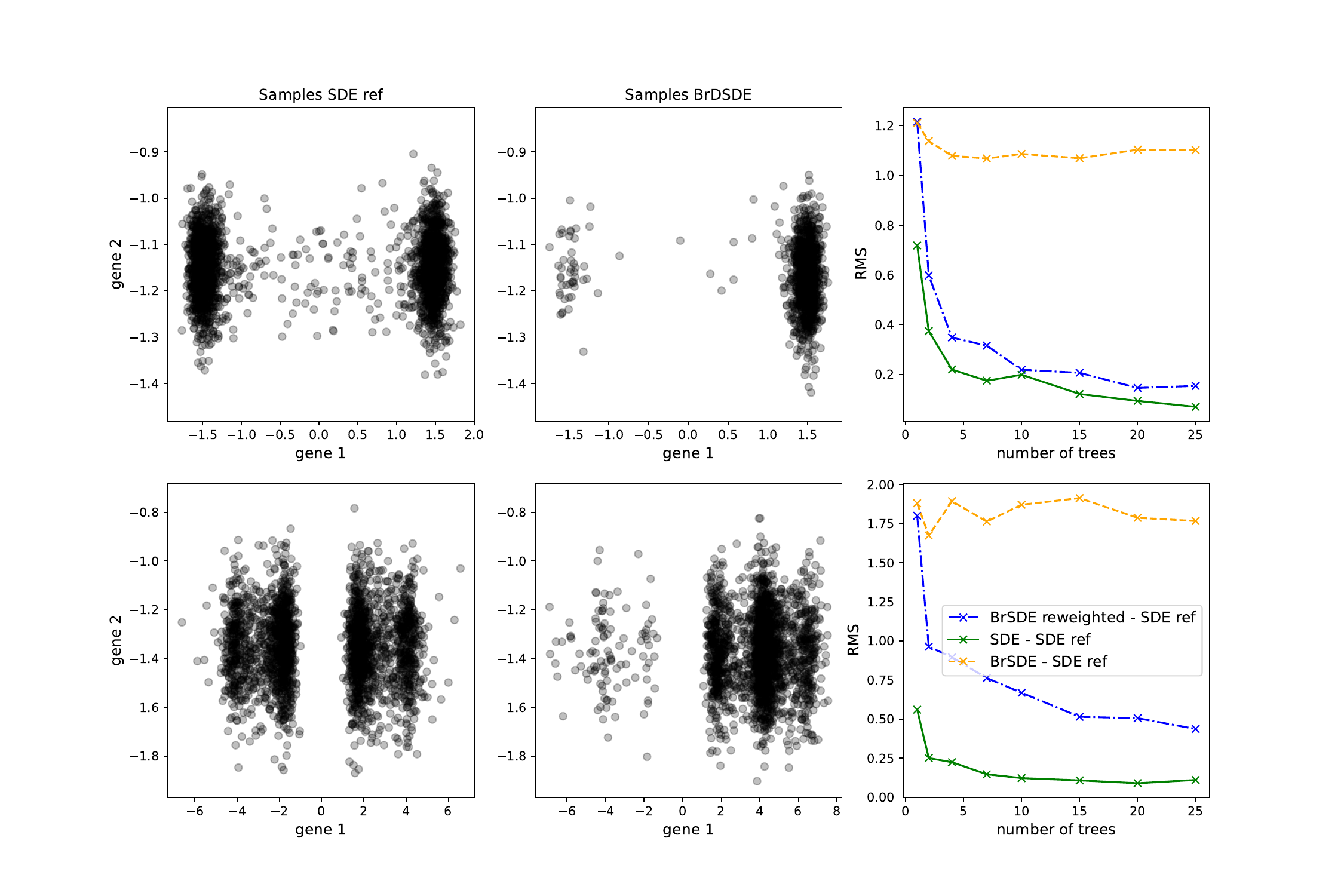}
\caption{Numerical convergence of reweighted empirical measure associated to a branching SDE to the ground-truth distribution of the underlying SDE. First column: ground-truth distributions, obtained by simulating $2500$ cells with the SDE. Second column: empirical measure obtained by simulating $25$ trees containing, in total, $1050$ leaves (first line) and $2584$ leaves (second line), with the corresponding branching SDE. Third column: evolution of the RMS distance between the ground-truth and (green) the reweighted empirical measure, (red) the empirical measure (without reweighting), and (blue) the empirical distribution obtained by simulating the non-branching SDE with the same number of cells as the number of leaves obtained with the branching SDE. Each row of plots correspond to one of the branching SDEs described in in Appendix \ref{appendix_parameters}.
\resolved{\af{Change colors in third column to avoid red/green. We could also do different linestyles.}\ev{Done}}
}
\label{Figure2}
\end{figure}

The convergence of the path measure in Theorem \ref{thm_consistency_1} strongly relies on the convergence of the reweighted empirical distributions to the temporal marginals of the SDE \eqref{eq_SDE}. Although a quantification of such convergence rate is beyond the scope of this article, in Fig.~\ref{Figure2} we investigate numerically how the marginals converge as the number $K$ of observed trees increases. To do so, we simulate the long-time behavior of two branching SDEs, described fully in Appendix \ref{appendix_parameters}. Our accuracy metric is the the root-mean-square (RMS) Energy Distance distance~\cite{szekely2013energy} between the reweighted empirical distribution and the ground-truth distribution without branching (the latter computed by simulating thousands of cells under the underlying SDE). We plot the evolution of this distance as a function of the number of simulated trees used to form reweighted empirical distribution. As a reference point, for each $K$ we simulate the same total number of cells using only the underlying SDE and we compare the RMS distance between the empirical distributions thus obtained and the ground-truth.

 Because the reweighted distribution uses information on $\mathcal P(\mathbb N \times \mathcal X)$, which is a much bigger space than $\mathcal P(\mathcal X)$, it is \textit{a priori} plausible that accurately reconstructing the marginals requires much more data than in the case without branching. However, the results in the third column of Fig.~\ref{Figure2} suggest that it is not the case: reweighting in these examples (green line), although somewhat worse than using data directly from the underlying SDE (blue line), significantly improves on not reweighting (orange line) as soon as $K \ge 2$. 
 
 For the second branching SDE, on the second row of Figure~\ref{Figure2}, the distance remains non-negligible out to 25 trees only because the branching rate is particularly high in a shallow well located around $x \approx 6$ in the first dimension. Thus, cells are likely to reach and stay in this well for the branching SDE but not for the underlying SDE. Reweighting by the generation numbers reduces the weight assigned to these cells, but requires more trees to fully adjust away the proliferation than are required for the branching SDE in the first line with only two wells.

The numerical evidence of Figure~\ref{Figure2} suggests that using the reweighted distribution for trajectory inference is quite powerful in that it allows us to solve a problem in a very big space corresponding to branching SDEs with lineage tracing, while requiring a comparable amount of data to the simpler problem of trajectory inference for SDEs, posed in a much smaller space.

\resolved{\af{I'm not sure what you mean by ``stable with respect to the law of large numbers''. Just that $\hat \rho(m)$ converges quickly to the intensity as $K$ grows? Are you arguing that $\mathbb N \times \mathcal X$ being bigger than $\mathcal X$ makes the problem easier or harder?}\ev{I've removed this term. Is that clear enough now? Don't hesitate to reformulate this part!}}

\subsection{Computational inference of trajectories and model characteristics}
\label{subsec_applications}

We are now going to show how Theorem \ref{thm_consistency_1} can be efficiently applied to reconstruct the drift and the birth rate of a branching SDE, with examples from simulated datasets with lineage tracing.
 
\subsubsection{Trajectories of the time-varying distribution}
\label{subsec_MFL}

We use the method described in Chizat et al.~\cite{zhang2022trajectory} for reconstructing the trajectories of a time-varying probabilistic distribution from a time series of its temporal marginals. This algorithm optimizes a close variant of the functional $F_{N,\lambda,h}$ used in the consistency result of Theorem \ref{thm_consistency_1}. Explicitly, Chizat et al. propose finding
\begin{align}
    \label{dynamical_MFL}
    \inf\limits_{R \in \mathcal P(\textrm{c\`adl\`ag}([0, T], \mathcal X))}\left\{\tau H(R | W^\tau) +  \frac{1}{\lambda}\sum_{i=1}^N \Delta t_i H\left(\hat \rho_{t_i} | \Phi_h * R_{t_i}\right)\right \},
\end{align}
where $(\hat \rho_{t_i})$ is the sequence of empirical distributions describing the data at every timepoint. Their algorithm is described by a Mean-Field Langevin dynamics~\cite{chizat2022mean}, and we refer to it as the \emph{MFL algorithm} in the rest of the paper. In practice, it provides a time-varying sum of $M$ Dirac point on $\mathcal X$, which converges to the path-measure corresponding to the unique solution of the problem \eqref{dynamical_MFL} as $M \to \infty$. As the reweighted empirical measures \eqref{reweighted_empirical_measures} are probabilistic distributions, we can directly apply the MFL method by, at every $t_i$, replacing the empirical distribution $\hat \rho_{t_i}$ by the corresponding reweighted empirical distribution. We call this method the \emph{reweighting method} in the following.

\bigskip

\begin{figure}
\centering
\includegraphics[trim=4.5cm 0cm 14cm 0cm, width=\textwidth]{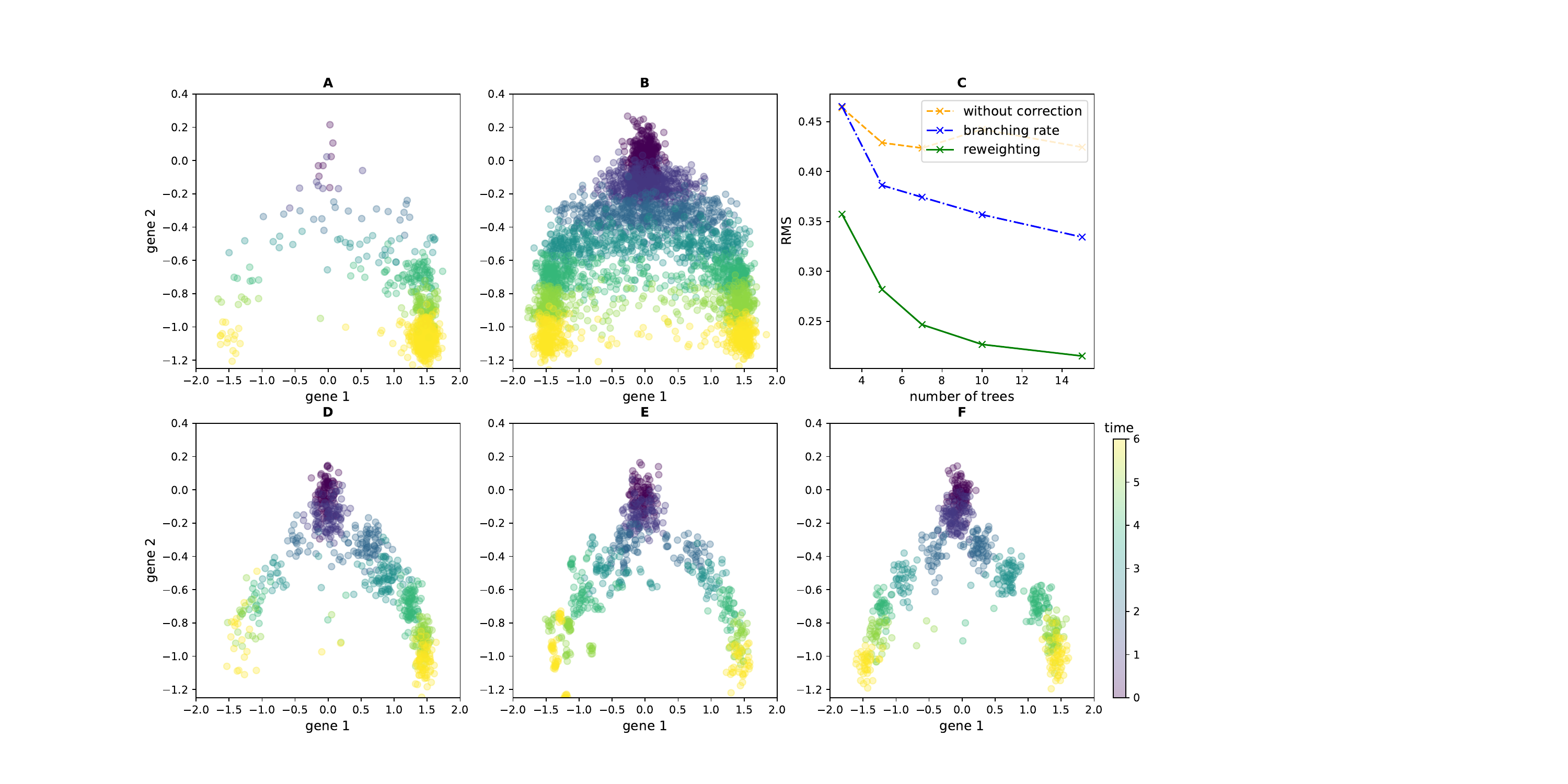}
\caption{The time-varying distribution of an SDE with a double-well potential can be reconstructed using data from its associated branching SDE. Panel A shows the observed data when 5 independent trees are generated using potential $V_1$ from Appendix \ref{appendix_parameters}. 
B shows the ground-truth distribution we aim to recover, simulated using the underlying SDE with 500 cells at each timepoint.
The cumulated RMS distance to this ground-truth at each timepoint is rather high if the MFL algorithm is applied with no correction for proliferation (C, dashed).
Using a heuristic correction for known growth rates reduces the error (C, dashdotted), and applying our reweighting method reduces it further (C, solid line).
The second row shows the reconstructed distributions using the three approaches: MFL without correction (D, visibly biased towards the more proliferative right well), MFL with the growth rate correction (E), and MFL on the reweighted marginals (F).
}
\label{Figure3}
\end{figure}

Figure \ref{Figure3} illustrates the results of applying the MFL algorithm to data from a branching SDE with two wells (detailed as potential $V_1$ in Appendix~\ref{appendix_parameters}). We compare three approaches: using no correction for proliferation (D), applying a heuristic correction (\cite{zhang2022trajectory}, Section 4) that requires the branching rate to be known (E), and our new reweighting method (F). Reweighting significantly improves on both other strategies, even though the heuristic correction is given ground-truth branching rates that are never known in practice.
The true branching rates are insufficient here they only take into account the \emph{average effect} of branching. In cases where sampling variation causes the ratio of cells in each well to differ from the expected average, using the branching rates directly biases the reconstruction. In contrast, our method uses the generation numbers to account for the exact effect of branching in each experimental realization. 

This efficient reconstruction of the time-varying distribution associated to the underlying potential demonstrates the utility of the reweighting method and, more generally, of the information about generation numbers. As shown in the next sections, the recovered trajectory can subsequently be used to estimate the drift and birth rates that define the underlying SDE.

\subsubsection{Estimation of the drift and branching rates}

Mean-field Langevin with our reweighting enables reconstruction of a most-likely time-varying distribution $p^*$ of the process characterized by the master equation \eqref{master_equation_onlydeath}. We are now going to use this result to detail how to estimate the parameters of the model, \emph{i.e} the branching rates and the drift of the process.

\bigskip

We recall that the core of the MFL algorithm consists in computing the Schr\"odinger potentials of the Schr\"odinger problem between each pair of timepoints $(t_i, t_{i+1})$. These potentials are updated at each step of the algorithm together with the marginals of time-varying distribution (see Proposition 3.2 from Chizat et al.~\cite{zhang2022trajectory}). We denote the potentials associated to the optimal marginals, obtained at the convergence, $\phi^*_{i, i+1}$ and $\psi^*_{i, i+1}$. It is well known that these potentials are directly related to the drift of the optimal process associated to these marginals. More precisely, in case where the marginals are continuous we have the following lemma:

\begin{lem}[see for example~\cite{pavon1991free}]
\label{lem_exp_martingale}
With the previous notation, for all $i = 1\cdots,N$, the time-varying distribution $p^*$ characterizing the solution of the Schr\"odinger problem with temporal marginals $p_{t_i}$ and $p_{t_{i+1}}$ is solution of the following PDE in $[t_i, t_{i+1}]$:
\begin{align}
\label{eq_master_eq_withoutb}
\partial_t p^{*} = \mathcal L^*_{-u, \tau}p^*,
\end{align}
where the function $u$ from $\,[t_i, t_{i+1}] \times \mathcal{X}$ to $\mathbb{R}^+$ satisfies
\begin{align*}
    \partial_t e^{\frac{u}{\tau}} = -\frac{\tau}{2} \Delta e^{\frac{u}{\tau}},
\end{align*}
for all $t\in[t_i, t_{i+1}]$, with initial condition $u(t_i) = -\phi^*_{i, i+1}$.
\end{lem}

This lemma thus provides a way of characterizing the drift $v^*$ between each pair of timepoints from the optimal temporal marginals obtained by the MFL algorithm and the associated Schr\"odinger potentials, with the formula $v^* = \nabla u$. A similar way of characterizing an optimal birth rate $b^*$ would be to consider the optimal time-varying intensity with branching, denoted $\rho^{*}$, which is solution in the weak sense of the equation:
\begin{align}
    \label{eq_master_eq_b}
    \partial_t \rho^{*} = \mathcal L^*_{-u, \tau} \rho^{*} + b^* \rho^{*}.
\end{align}
Indeed, provided that we have access to the optimal temporal marginals of $\rho^{*}$, we could then approximate:
\begin{align}
\label{reconstruction_proliferation}
   \forall i=1,\cdots,T,\,\forall x \in \mathcal X:\, b^*(t_i,x) \approx \frac{1}{t_{i+1} - t_i} \ln \left(\mathbf E^{P^*}\left[\frac{\rho^{*}_{t_{i+1}}(X_{t_{i+1}})}{\rho^{*}_{t_i}(X_{t_{i}})} \bigg| X_{t_i} = x\right]\right),
\end{align}
where the expected value is taken under the optimal coupling $P^* = e^{\frac{\phi^*_{i, i+1}+\psi^*_{i, i+1}}{\tau}}W^{\tau}_{t_i, t_{i+1}}$\resolved{\af{It feels odd to me to use $W^\tau$ for both the Brownian motion stochastic process and the coupling the process entails.}\ev{Should be solved}}. Considering that the optimal temporal marginals of $\rho^{*}$ simply correspond to the sequence of observed empirical intensities $(\hat \mu_{t_i})_{i=1\cdots,n}$, we could estimate $b^*$ at each timepoint with formula \eqref{reconstruction_proliferation}.

\bigskip

However, in practice we have only discrete measures and these estimations of $v^*$ and $b^*$ using the Schr\"odinger potentials are not directly applicable. Indeed, the formula $v^* = \nabla u$ can be difficult to use when the gene expression space $\mathcal X$ is high-dimensional, as the gradient is not accessible in most of the directions. Moreover, the optimal birth rate $b^*$ defined by \eqref{reconstruction_proliferation} is not directly computable when the optimal path-measure $P^*$ is computed with the MFL algorithm, since its support does not correspond with the support of the observed empirical intensities. We need alternative strategies to find both $v^*$ and $b^*$

\bigskip

For $v^*$, we follow the proposal in Lavenant et al.~\cite{Lavenant2021} to use the approximation
\begin{align}
    \label{reconstruction_velocity}
    v(t_i, x) = \mathbf E^{P^*}\left[\frac{X_{t_{i+1}} - X_{t_i}}{t_{i+1} - t_i} \bigg| X_{t_i} = x \right],
\end{align}
which becomes exact in the limit $t_{i+1} - t_i \to 0$.

\bigskip

For $b^*$, in order to make the formula \eqref{reconstruction_proliferation} usable, we estimate for each timepoint $t_i$ a most-likely marginal intensity $\rho^{*}_{t_i}$ having the same support as $p^*_{t_i}$. We start from the reweighted empirical distribution $\hat{\rho}_{t_i} = \sum\limits_m \frac{1}{2^m} \hat \mu_{t_i}(m)$, and propose the following two-step algorithm:
\begin{enumerate}
    \item Find the optimal coupling $\pi^*$ between the cells characterizing the two experimental distributions $p^*_{t_i}$ and $\hat{\rho}_{t_i}$, by solving the optimal transport problem $W^2(p^*_{t_i}, \hat{\bar \rho}_{t_i})$;
    \item For every cell $x\in \textrm{Supp } p^*_{t_i}$, we compute an associated $m(x)$:
    $$m(x) = \frac{\sum\limits_{(m, y) \in \textrm{Supp } \hat \mu_{t_i}} m \pi^*(x,y)}{\sum\limits_{y \in \textrm{Supp } \hat \mu_{t_i}}\pi^*(x,y)}.$$
    We then consider:
        \begin{align}
        \label{eq_transfo_rho*_rhob*}
            \rho^{*}_{t_i}(x) = 2^{m(x)}p^{*}_{t_i}(x).
        \end{align}
\end{enumerate}
As an intuitive justification, it is easy to verify that if we do not use the MFL algorithm and simply set $\rho^{*}_{t_i}(\cdot) = \sum\limits_m \frac{1}{2^m} \hat \mu_{t_i}(m, \cdot)$, then $\pi^*(x,y) = \delta_x(y)$ and we recover $\rho^{*}_{t_i} = \sum\limits_m \hat \mu_{t_i}(m, \cdot)$. 

\bigskip

We apply these methods on the trajectories inferred in Fig.~\ref{Figure3}, and compare in Fig.~\ref{Figure4} the estimated velocity field and birth rates with the simulation's ground truth described in Appendix \ref{appendix_parameters}.

\begin{figure}
\includegraphics[width=1.3\textwidth]{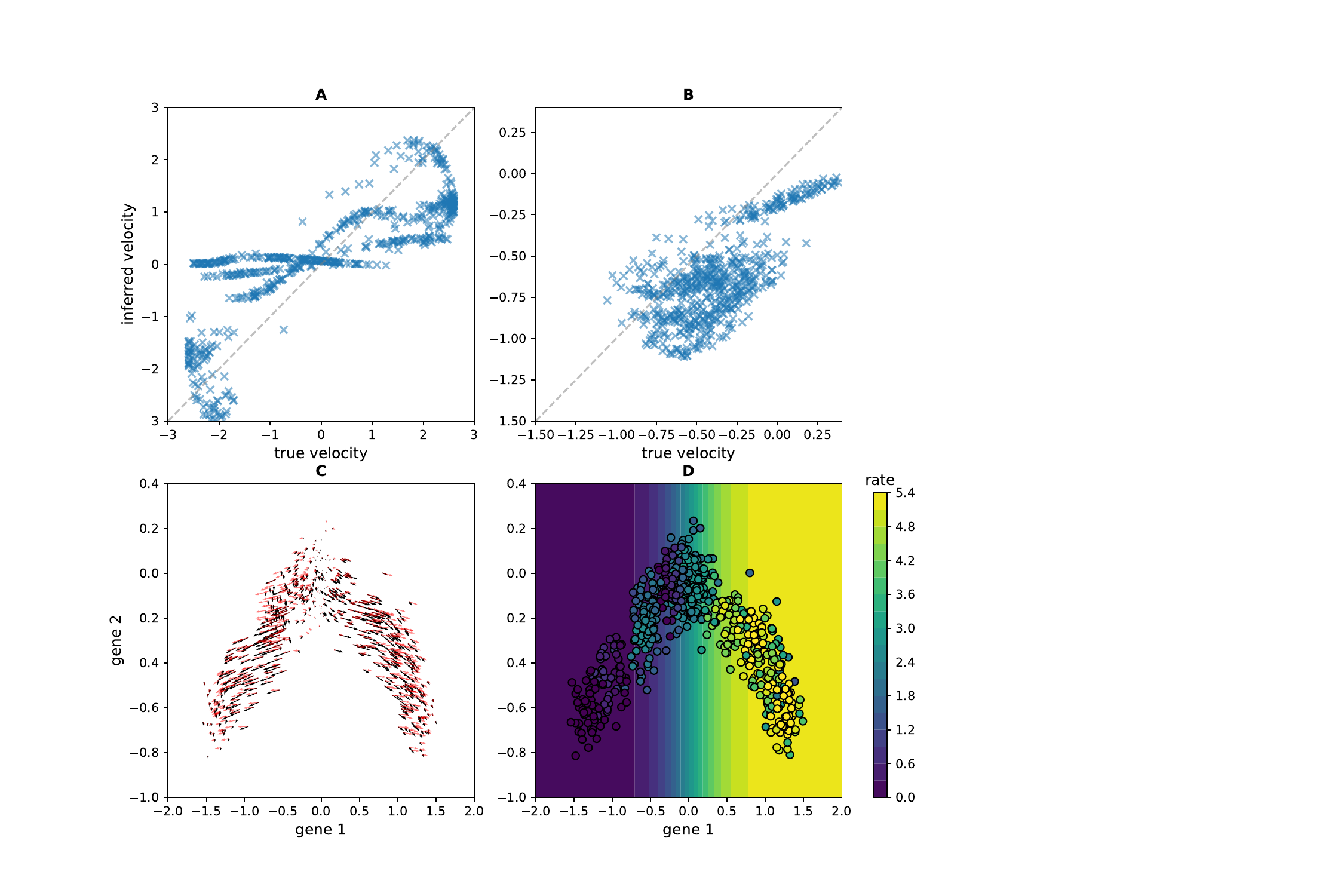}
\caption{Estimation of (A)-(C) the velocity fields and (D) the birth rates reconstructed using the methods detailed in Section \ref{subsec_applications}. For the velocity fields, we represent in (A) (resp. B) the comparison between the inferred velocity field, with the formula \eqref{reconstruction_velocity} applied to the time-varying distribution reconstructed with the MFL algorithm \eqref{reconstruction_velocity}, and the ground-truth using the parameters of the first branching SDE described in Appendix \ref{appendix_parameters}, for the first gene (resp. the second gene). The grey line represent the diagonal which would correspond to a perfect fit. In (C) we plot these two velocity fields along the trajectory, in dark for the inferred one and in red for the ground-truth. For the birth rate, we first use \eqref{eq_transfo_rho*_rhob*} to estimate from this distribution the new distribution with branching $\rho^{*}$; second, we compute the associated birth rate using \eqref{reconstruction_proliferation}. The background of (D) \resolved{\af{Consider matching the opacity in the background and the points and adding white or black circles around the points so they're still visible.}} corresponds to the true birth rate used for the simulation. The data corresponds to the reconstructed trajectories of $100$ cells at each timepoints with the MFL algorithm from time-series of snapshots obtained by simulating $15$ trees.\resolved{\af{Should we have some quantitative accuracy metric?}\ev{We decided that no, as the results are pretty clear}
\af{Could we show the ground truth velocities we're trying to recover?}\ev{Done}}}
\resolved{\af{The velocity plot is very informative if I enlarge it substantially, but difficult to parse at a glance. Is there another way we could plot the velocities? Perhaps two scatter plots, one for true vs. recovered velocity in Gene 1 and one for true vs. recovered velocity in Gene 2 (and leaving the current (A) to show accuracy as a function of position)?}
\af{What are the sampling times?}
\af{It looks like the method is particularly struggling to recover the Gene 2 velocity in regions where not many cells are sampled at a time point. The mismatch at the bifurcation point in the middle is prominent in the plot disproportionately to how important it seems to me. Overall, (A) feels like it's highlighting mistakes over places where the velocity is correctly recovered. Two changes that might (or might not) help: plotting the recovered velocities on top rather than on bottom (so the wells at the bottom are more black than red) and changing the opacity so overlapping vectors are less obscured.}
\af{It might also help if the potential weren't time-dependent, so that the true velocity vectors didn't cross each other in a hard-to-parse way.}
\af{Sorry for so many comments - I think the method is doing well and I'm struggling to figure out how to make that come across in the figure.}}
\label{Figure4}
\end{figure}

\section{Trajectory inference with death and subsampling}
\label{sec_method_bias}

When cells can die, the generation number $\tilde m(X)$ observed with lineage tracing for a cell $X$ at time $t$ may differ from the real generation number $m(X)$, because $\tilde m(X)$ only takes into account divisions for which both branches have a descendant alive at $t$ (Figure~\ref{Figure1}).
The reasoning of Section~\ref{sec_consistencytheorem1} is no longer valid.
Subsampling causes an identical problem: if a division does not result in two branches that are subsampled at $t_i$, then it cannot be taken into account in $\tilde m(X)$. 
We consider these two situations together in this section. The case with only death or the case with only subsampling corresponds to setting $q(t_i) = 1$ or $d(t,x)=0$, respectively, in the results that follow.

\bigskip

The main analytical obstacle in the case with death is that the branching property~\eqref{branching_property} does not hold for the process on $(\tilde m, x)$. In particular, the death of cell $x$ not only removes $x$ but also changes $\tilde m$ for every cell descended from the parent of $x$. Those interactions between cells introduce non-local terms in the master equation characterizing the evolution of the intensity of the process, significantly complicating any calculations. Because that complexity, we restrict the goals of this section to the following questions:

\begin{enumerate}
    \item (Section~\ref{subsec_bias_analysis}) What bias results from applying the method in Section~\ref{sec_consistencytheorem1}, \textit{i.e} finding the probabilistic distribution minimizing a functional of the form \eqref{functional_FTLH} where the empirical distributions corresponds to the reweighted empirical distributions of $\tilde Z_t$?
    \item (Section~\ref{subsec_bias_reduction_method}) How can the information from \emph{lineage tracing}, not restricted to $\tilde m(X)$, be used to partially remove this bias?
    \item (Section~\ref{subsec_estimation_subsampling_rate}) Under what experimental conditions is the bias likely to be small?
\end{enumerate}

\bigskip

Strictly speaking, throughout the paper we should give cells labels to refer to them, because many cells could be in the same position at the same time. However, when the number of initial cells is finite, Assumption \eqref{assumption_rates} implies that almost surely each cell at one time has a unique state. For the sake of simplicity, we therefore identify each cell at a given time by its position in $\mathcal X$.

\subsection{Bias analysis}
\label{subsec_bias_analysis}

Using the same notation as in the previous sections for the process with real generation number $m$, we now write $\tilde Z_t(\tilde m)$ for the random variable describing a collection of tuples in $\mathbb N \times \mathcal X$ at time $t$ observed after subsampling and let $\mathcal{\tilde N}(t) := \textrm{Supp}\,\tilde Z_t$. The law of $\tilde Z_t$ is described by a path-measure $\tilde P \in \mathcal P(\textrm{c\`adl\`ag}([0, T], \mathcal M_+(\mathcal X)^{\mathbb N}))$.
In this setting, $\tilde P$ does not satisfy the branching property defined above but a weaker property where we condition on each cell having at least one observed descendant: 

\begin{itemize}
    \item \textbf{The conditional branching property}: Let $t_i \in [0,T]$ and $t \leq t_i$. For any $x \in \mathcal X$, we denote $\mathcal{\tilde N}_{x}(t) := \textrm{Supp}\,\tilde Z_t^{x}$, where $\tilde Z_t^{x}$ is the random variable describing the collection of tuples at time $t$ descending from the same common ancestor in $x$ at time $0$. For any sum of Dirac masses $\mu$ in $\mathbb N \times \mathcal X$ and any family of measurable functions $\{u(m)\}_{m\in \mathbb N}$ from $\mathcal X$ to $\mathbb R^+ \cup \{+\infty\}$, we have:
    \begin{align}
    \label{branching_property_conditional}
    \mathbf E^{\tilde P}_{\mu}\bigg[\prod\limits_{(\tilde M,X) \in \mathcal{\tilde N}(t)}&u(\tilde M, X) \bigg|\, \forall (\tilde m,x) \in \textrm{Supp}\,\mu,\,\big|\mathcal{\tilde N}_{ x}(t_i)| \geq 1 \bigg] = 
    \\&\prod\limits_{(\tilde m,x) \in \textrm{Supp}\,\mu} \mathbf E^{\tilde P}_{(\tilde m,x)}\bigg[\prod\limits_{(\tilde M, X) \in \mathcal{\tilde N}(t)}u(\tilde M, X)\bigg|\, |\mathcal{\tilde N}(t_i)| \geq 1\bigg]. \notag
    \end{align}
\end{itemize}

This equality is immediate for the branching SDE without generation numbers because of the independence of branches stated by the branching property. For the branching SDE with observable generation numbers, each branch starting from a cell in the initial distribution evolves in $\mathbb N \times \mathcal X$ independently from the others conditional on the fact that all these branches survive, \textit{i.e} that each cell in the initial distribution has at least one subsampled descendant. 
Indeed, the only event in a branch that can impact the other branches is its extinction, which would update the observable generation numbers of cells in other branches. The initial branches are therefore independent conditional on none of them going extinct, as expressed by \eqref{branching_property_conditional}.

\bigskip

Now, for every timepoint $t_i \in [0,T]$, we consider a new family of random variables $(\tilde Z_t^{t_i}(\tilde m))_{m \in \mathbb N}$, describing at every earlier time $t \in [0, t_i]$ the collection of tuples that have at least one subsampled descendant at time $t_i$. We denote $(\mathcal{\tilde F}^{t_i}_t)_{t \in [0, t_i]}$ the natural filtration associated to this stochastic process. The conditional branching property \eqref{branching_property_conditional} holds for this new process as well.

For each family, we define the family of time-varying intensities $\{\rho^{t_i}(\tilde m)\}_{m\in \mathbb N}$, by:
\begin{align}
\label{def_intensities_conditional}
    \forall \tilde m \geq 0,\,\forall t \leq t_i:\, \rho_t^{t_i}(\tilde m):\, A \to \mathbf E^{\tilde P}\big[\tilde Z_t^{t_i}(\tilde m,A)\big|\, |\mathcal{\tilde N}(t_i)| \geq 1 \big].
\end{align}

Importantly, at each measurement time $t_i$, the new random variables coincide with our observations: $\tilde Z_{t_i}^{t_i}(\tilde m) = \tilde Z_{t_i}(\tilde m)$. This means that the family of intensities $\{\rho^{t_i}(\tilde m)\}$ coincides at time $t_i$ with the family of intensities associated to the temporal marginals of the path-measure $\tilde P_{t_i}$ that describes the branching SDE with observable generation numbers, conditional on subsampling at least one cell at time $t_i$. 

\bigskip

The importance of conditioning on descendants being subsampled leads us to define a branch survival probability:
\begin{align}
        \label{assumption_reg_survival}
    S^{t_i}:\,(t,x)\to \mathbf P_x(\mathcal{\tilde N}(t_i-t) \geq 1).
\end{align}
We can now state the main result of this section:
\begin{prop}
\label{prop_caracteristion_rhoti}
    Let $t_i \in [0, T]$. Under Assumption~\eqref{assumption_rates} and assuming that the branch survival probability $S^{t_i}(\cdot, x)$ is continuous for all $x \in \mathcal X$ and $S^{t_i}(t, \cdot) \in C^1(\mathcal X)$ for all $t \in [0,t_i]$, the family $\{\rho^{t_i}(\tilde m)\}_{\tilde m\in \mathbb N}$ defined by \eqref{def_intensities_conditional} solves the following system of PDEs in the weak sense:
\begin{align}
\label{system_master_equation_conditional}
\forall \tilde m \geq 0:\, \partial_t \rho^{t_i}(\tilde m) = \mathcal L^{*}_{\Psi - \tau \ln S^{t_i}, \tau}\rho^{t_i}(\tilde m) +  2bS^{t_i} \rho^{t_i}(\tilde m-1) - b S^{t_i}\rho^{t_i}(\tilde m),
\end{align}
with initial condition $\rho^{t_i}_0(\tilde m, \cdot) \propto {\mu_0 S^{t_i}(0, \cdot)\mathds 1_{\tilde m = 0}}$.
\end{prop}

Note that this system of PDE coincides with Eq.~\ref{master_equation_system} for new branching SDE with a modified potential $\Psi - \tau \ln S^{t_i}$, a modified birth rate $bS^{t_i}$, and no death. In the following, we will call the quantity
\begin{align}
\label{def_bias_potential}
\Psi^{t_i} := -\tau \ln S^{t_i},
\end{align}
the \emph{bias} in the potential; $-\nabla \Psi^{t_i} = \tau \nabla \ln S^{t_i}$ is then the bias in the drift.

\begin{proof}
The proof of this proposition follows the same steps as the proof of \eqref{master_equation_system}, using the Markov property, invariance by translation, and the  conditional branching property. In particular, by the conditional branching property, it is enough to prove the proposition starting from a cell in an arbitrary tuple $(\tilde m,x) \in \mathbb N \times \mathcal X$. At a small time $\delta t$, the set of complementary events that we consider are now:
\begin{itemize}
    \item $A^b_2(\delta t)$: the cell has divided into two cells and the two cells each have a surviving descendant at $T$;
    \item $A^b_1(\delta t)$: the cell has divided into two cells and only one cell has a surviving descendant at $T$;
    \item $A^b_0(\delta(t)$: the cell has divided into two cells and no cell has a surviving descendant at $T$;
    \item $A^d(\delta t)$: the cell is dead;
    \item $A^c(\delta t)$: the cell has neither divided nor died.
\end{itemize}
We denote $u(t_i, t, \tilde m, x) = \mathbf E_{(\tilde m, x)}\left[\prod\limits_{\substack{(\tilde M, X) \in \mathcal{\tilde N}(\delta t) \\ \mathcal{\tilde N}_{X}(t_i-t) \geq 1}} u_0(\tilde M,X) \bigg| |\mathcal{\tilde N}(t_i)| \geq 1\right]$, where the expected value is taken under the law of $\tilde P$. It is clear that the probability of $A^b_0$ and $A^d$ are $0$ conditional on $\{|\mathcal{\tilde N}(t_i)| \geq 1\}$, and that the probability of $\{|\mathcal{\tilde N}(t_i)| \geq 1\}$ is $1$ conditionally to $A^b_2(\delta t)$ and $A^b_1(\delta t)$. We have then:
\begin{align*}
u(t_i, \delta t, \tilde m, x) &= (1 - \mathbf P_{x}(A^b_2(\delta t) \cup A^b_1(\delta t)\big| |\mathcal{\tilde N}(t_i)| \geq 1)) \times \\ &\hspace{4.79cm} \mathbf E_{(\tilde m, x)}\left[\prod\limits_{\substack{(\tilde M, X) \in \mathcal{\tilde N}(\delta t) \\ \mathcal{\tilde N}_{X}(t_i-t) \geq 1}} u_0(\tilde M,X) \bigg| (|\mathcal{\tilde N}(t_i)| \geq 1) \cap A^c\right] \\ &+ \mathbf P_{x}(A^b_2(\delta t)\big| |\mathcal{\tilde N}(t_i)| \geq 1)\times \mathbf E_{(\tilde m, x)}\left[\prod\limits_{\substack{(\tilde M, X) \in \mathcal{\tilde N}(\delta t) \\ \mathcal{\tilde N}_{X}(t_i-t) \geq 1}} u_0(\tilde M,X) \bigg| A^b_2(\delta t) \right] \\ &+ \mathbf P_{x}(A^b_1(\delta t)\big| |\mathcal{\tilde N}(t_i)| \geq 1)\times \mathbf E_{(\tilde m, x)}\left[\prod\limits_{\substack{(\tilde M, X) \in \mathcal{\tilde N}(\delta t) \\ \mathcal{\tilde N}_{X}(t_i-t) \geq 1}} u_0(\tilde M,X) \bigg| A^b_1(\delta t)\right].
\end{align*}
Moreover, using Bayes' law and a little algebra, we have:
\begin{align*}
    &\mathbf P_{x}(A^b_2(\delta t) \cup A^b_1(\delta t)\big| |\mathcal{\tilde N}(t_i)| \geq 1) = \mathbf P_{x}(A^b_2(\delta t) \cup A^b_1(\delta t)) \times  \frac{\mathbf P_{x}(|\mathcal{\tilde N}(t_i)| \geq 1\big| A^b_2(\delta t) \cup A^b_1(\delta t))}{\mathbf P_{x}(|\mathcal{\tilde N}(t_i)| \geq 1)},\\
    &\mathbf P_{x}(A^b_2(\delta t)\big| |\mathcal{\tilde N}(t_i)| \geq 1) = \mathbf P_{x}(A^b_2(\delta t) \cup A^b_1(\delta t)\big| |\mathcal{\tilde N}(t_i)| \geq 1) \times \frac{\mathbf P_{x}(|\mathcal{\tilde N}(t_i - \delta t)| \geq 1)^2}{\mathbf P_{x}(|\mathcal{\tilde N}(t_i - \delta t)| \geq 1\big| A^b_2(\delta t) \cup A^b_1(\delta t))},\\
   &\mathbf P_{x}(A^b_1(\delta t)\big| |\mathcal{\tilde N}(t_i)| \geq 1) = \mathbf P_{x}(A^b_2(\delta t) \cup A^b_1(\delta t)\big| |\mathcal{\tilde N}(t_i)| \geq 1) \\ &\hspace{7.5cm} \times \frac{2\mathbf P_{x}(|\mathcal{\tilde N}(t_i - \delta t)| \geq 1) (1-\mathbf P_{x}(|\mathcal{\tilde N}(t_i - \delta t)| \geq 1))}{\mathbf P_{x}(|\mathcal{\tilde N}(t_i - \delta t)| \geq 1\big| A^b_2(\delta t) \cup A^b_1(\delta t))}.
\end{align*}
Note that these quantities are well defined since under Assumption \eqref{assumption_rates} the function $S^{t_i}(t, \cdot)$ is strictly positive for all $t$.
We then obtain by continuity of $S^{t_i}(\cdot, x)$:
\begin{align*}
    &\lim\limits_{\delta t \to 0}\frac{\mathbf P_{x}(A^b_2(\delta t) \cup A^b_1(\delta t)\big| |\mathcal{\tilde N}(t_i)| \geq 1)}{\delta t} = b (2 - S^{t_i}(0,x)),\\
    &\lim\limits_{\delta t \to 0}\frac{\mathbf P_{x}(A^b_2(\delta t)\big| |\mathcal{\tilde N}(t_i)| \geq 1)}{\delta t} = b S^{t_i}(0,x),\\
    &\lim\limits_{\delta t \to 0}\frac{\mathbf P_{x}(A^b_1(\delta t)\big| |\mathcal{\tilde N}(t_i)| \geq 1)}{\delta t} = 2b (1 - S^{t_i}(0,x)).
\end{align*}

Finally, thanks to the conditional branching property and following the same reasoning as in the proof of \eqref{master_equation_system}, Step 1.:
\begin{align*}
\lim\limits_{\delta t \to 0} \frac {u(t_i, \delta t, \tilde m, x) - u_0(\tilde m, x)}{\delta t} = &\lim\limits_{\delta t \to 0} \frac {\mathbf E_x^p\left[u_0(\tilde m, X_{\delta t})\bigg| |\mathcal{\tilde N}(t_i)| \geq 1\right] - u_0(\tilde m, x)}{\delta t} \\ &- b S^{t_i}(0,x) u_0(\tilde m, x) + b S^{t_i}(0,x) u_0(\tilde m+1, x)^2.
\end{align*}
To find the value of the right-hand side term in the first line, we use the It\^o formula, which states that for all $\tilde m,x$:
$$u_0(\tilde m, X_{\delta t}) = u_0(\tilde m, x) + \int_0^{\delta t} \mathcal L_{\Psi, \tau} u_0(\tilde m, X_s)\mathrm ds + \tau \int_0^{\delta t} \nabla u_0(\tilde m, X_s)\mathrm dB_s,$$
ensuring that, as long as $u_0 \in C^2_b(\mathcal X)$:
\begin{align*}
\lim\limits_{\delta t \to 0} \frac {\mathbf E_x^p\left[u_0(\tilde m, X_{\delta t})\bigg| |\mathcal{\tilde N}(t_i)| \geq 1\right] - u_0(\tilde m, x)}{\delta t} = &\mathcal L_{\Psi, \tau} u_0(\tilde m, x) + \\&\tau \sum\limits_{j=1}^g \partial_{x_j} u_0(\tilde m, x) \lim\limits_{\delta t \to 0} \frac{1}{\delta t}\mathbf E_{x}^p\left[B_{\delta t}^j - x\bigg| |\mathcal{\tilde N}(t_i)| \geq 1\right].
\end{align*}

Using Bayes' rule once again, we have in every direction $j = 1,\cdots,g$:
\begin{align}
\notag
\lim\limits_{\delta t \to 0} \frac{1}{\delta t}\mathbf E_{x}^p\left[B_{\delta t}^j - x\bigg| |\mathcal{\tilde N}(t_i)| \geq 1\right] = &\lim\limits_{\delta t \to 0} \mathbf E_{\mathcal{\tilde N}(0, \frac{1}{\delta t})} \left[{Y}\frac{\mathbf P_{x + \delta tYe_j}(|\mathcal{\tilde N}(t_i - \delta t)| \geq 1)}{\mathbf P_x(|\mathcal{\tilde N}(t_i)| \geq 1)} \right],\\
= &\lim\limits_{\delta t \to 0} \mathbf E_{\mathcal{\tilde N}(0, \frac{1}{\delta t})} \left[{Y}\frac{S^{t_i}(\delta t, x + \delta t Y e_j)}{S^{t_i}(0, x)} \right].
\label{lim_gradient}
\end{align}

Using the fact that $S^{t_i}$ is differentiable w.r.t $x$, we can use its Taylor expansion in every direction $j=1,\cdots,g$ to obtain:

\begin{align*}
\mathbf E_{\mathcal{\tilde N}(0, \frac{1}{\delta t})} \left[{Y}\frac{S^{t_i}(\delta t, x + \delta t Y e_j)}{S^{t_i}(0, x)} \right] &= \mathbf E_{\mathcal{\tilde N}(0, \frac{1}{\delta t})} \left[{Y} \frac{S^{t_i}(\delta t, x)}{S^{t_i}(0, x)} + Y^2\delta t \frac{\partial_{x_j}S^{t_i}(\delta t, x)}{S^{t_i}(0, x)} + O(Y^3 \delta t^2)\right]
 \\
 & = \frac{\partial_{x_j}S^{t_i}(\delta t, x)}{S^{t_i}(0, x)} + O(\delta t).
\end{align*}
The remainder is $O(Y^4\delta t^3) = O(\delta t)$ because the symmetry of the distribution of $Y$ makes the odd terms in the Taylor expansion vanish.
By continuity of $S^{t_i}$ w.r.t $t$, the limit \eqref{lim_gradient} is then equal to $\partial_{x_j} \ln S^{t_i}(0, x)$.

Thus we obtain that $\forall \tilde m \in \mathbb N, \forall x \in \mathcal X$: 
\begin{align*}
\frac{\mathrm d}{\mathrm d t}u(t_i, t, \tilde m, x)\big|_{t = 0} &= \mathcal L_{\Psi + \Psi^{t_i}, \tau}  u_0(\tilde m,x) +  bS^{t_i} u_0(\tilde m+1, x)^2 - bS^{t_i} u_0(\tilde m, x),\\ &:= \mathcal{\tilde K}_{\Psi, \tau} u_0.
\end{align*}

\bigskip

We can now reason in a way very similar to Step 2. of the proof of \eqref{master_equation_system}, using the Markov property and the conditional branching property. Indeed, we have for all $t, s > 0$ such that $t+s \leq t_i$:
\begin{align*}
u&(t+s, \tilde m, x) = \mathbf E_{(\tilde m, x)}\left[\mathbf E_{(\tilde m, x)}\left[\prod\limits_{\substack{(\tilde M, X) \in \mathcal{\tilde N}(t+s) \\ \mathcal{\tilde N}_{X}(t_i-t-s) \geq 1}} u_0(\tilde M,X) \big| \left(|\mathcal{\tilde N}(t_i)| \geq 1\right) \cap \mathcal{\tilde F}^{t_i}_s \right] \Bigg| |\mathcal{\tilde N}(t_i)| \geq 1\right], \\
&= \mathbf E_{(\tilde m, x)}\left[\mathbf E_{\tilde Z_s^{t_i}}\left[\prod\limits_{\substack{(\tilde M,X) \in \mathcal{\tilde N}(t) \\ \mathcal{\tilde N}_{X}(t_i - t-s) \geq 1}}u_0(\tilde M, X) \bigg|\, \forall (\tilde M',X') \in \textrm{Supp}\,\tilde Z_s^{t_i},\,\big|\mathcal{\tilde N}_{X'}(t_i-s)| \geq 1\right] \Bigg| |\mathcal{\tilde N}(t_i)| \geq 1\right], \\ 
&= \mathbf E_{(\tilde m, x)}\left[\prod\limits_{\substack{(\tilde M', X') \in \mathcal{\tilde N}(s) \\ \mathcal{\tilde N}_{X'}(t_i-s) \geq 1}} \mathbf E_{(\tilde M,X)}\left[\prod\limits_{\substack{(\tilde M, X) \in \mathcal{\tilde N}(t) \\ \mathcal{\tilde N}_{X'}(t_i - t-s) \geq 1}}u_0(\tilde M, X)\bigg|\, |\mathcal{\tilde N}(t_i-s)| \geq 1\right]\Bigg| |\mathcal{\tilde N}(t_i)| \geq 1 \right],\\
&= \mathbf E_{(\tilde m, x)}\left[\prod\limits_{\substack{(\tilde M', X') \in \mathcal{\tilde N}(s) \\ \mathcal{\tilde N}_{X'}(t_i-s) \geq 1}} u(t_i - s, t, \tilde M', X')\bigg| |\mathcal{\tilde N}(t_i)| \geq 1 \right].
\end{align*}
Thanks to Step 1, we obtain the following system of PDEs:
\begin{align*}
\forall \tilde m \in \mathbb N, \forall x \in \mathcal X:\, \frac{\mathrm d}{\mathrm d s}u(t_i, t+s, \tilde m, x)\big|_{s = 0} = \mathcal{\tilde K}_{\Psi, \tau} u(t_i, t, \tilde m, x).
\end{align*}
Finally, Step 3. of the proof of Proposition \ref{prop_master_equation_realm} can be repeated without any change, and the initial condition follows directly from the definition of $\rho^{t_i}$, in \eqref{def_intensities_conditional}, using Bayes' rule.
\end{proof}

%\begin{rem}
%It is worth noticing that we could have defined equivalently the family of time-varying intensities $\{\rho^{t_i}(\tilde m)\}_{m\in \mathbb N}$, by:
%\begin{align*}
    %\forall \tilde m \geq 0,\,\forall t \leq t_i:\, \rho_t^{t_i}(\tilde m):\, A \to \mathbf E^{\tilde P}\big[\tilde Z_t^{t_i}(\tilde m,A) \big],
%\end{align*}
%and obtained that they solve, in the weak sense, the same system of master equations \eqref{system_master_equation_conditional}, the only difference being the mass of initial condition would be smaller than one, that is $\rho^{t_i}_0(\tilde m, \cdot) = \mu_0 S^{t_i}(0, \cdot)\mathds 1_{\tilde m = 0}$.
%\end{rem}

The final piece we need in order to understand the bias from applying the reweighting method in the presence of death and subsampling is to rule out the possibility that the process dies out entirely: 
\begin{conj}
\label{assumption_survival} 
We assume that the probability of subsampling at least one cell at any timepoint $\{t_1,\cdots,t_N\}$ can be approximated by one, \emph{i.e} that it exists $\varepsilon \ll 1$ such that for any test function $\theta$ on $\mathcal X$ and any timepoint $t_i$:
$$\forall t \in [0, t_i], \forall \tilde m \in \mathbb N:\, 
    \big|\mathbf E^{\tilde P}\left[\langle \theta, \tilde Z_t(\tilde m)\rangle\big| |\mathcal{\tilde N}(t_i)| \geq 1\right] - \mathbf E^{\tilde P}\left[\langle \theta, \tilde Z_t(\tilde m)\rangle\right] \big| \leq \varepsilon.
$$
\end{conj}

\begin{rem}
\label{rem_uptoeps}
A straightforward consequence of Assumption \ref{assumption_survival} is that, as $\tilde Z_{t_i}^{t_i}(\tilde m) = \tilde Z_{t_i}(\tilde m)$, the family of marginal intensities $\{\rho^{t_i}_{t_i}(\tilde m)\}_{\tilde m\in \mathbb N}$ \emph{coincides up to $\varepsilon$}  with the family $\{\rho_{t_i}(\tilde m)\}_{\tilde m\in \mathbb N}$, that is for all $\tilde m$ and all Borel sets $A \subset \mathcal X$:
$$|\rho^{t_i}_{t_i}(\tilde m,A) - \rho_{t_i}(\tilde m,A)| \leq \varepsilon.$$
\end{rem}

Assumption~\ref{assumption_survival} matches our biological setting perfectly for two reasons. First, clearly any data about living cells comes from a time when some cells were alive; second, typical measurements are done on growing tissues where the birth rate is higher than the death rate and extinction is unlikely. We deduce the following corollary, which is a natural extension of Corollary \ref{cor_realm} for the case with death and subsampling:

\begin{cor}
\label{cor_conditionalm}
Let $\bar{\rho}_{t_i} := \sum\limits_{\tilde m \geq 0} \frac{1}{2^{\tilde m}} \rho_{t_i}(\tilde m)$. Under Assumption \ref{assumption_survival} in addition to the assumptions of Proposition \ref{prop_caracteristion_rhoti}, at time $t_i$ $\bar \rho_{t_i}$ coincides up to $2 \varepsilon$ with the weak solution of the following PDE:
\begin{align}
\label{SDE_withbias}
\partial_t \rho = \mathcal L^*_{\Psi + \Psi^{t_i}, \tau} \rho,
\end{align}
where $\Psi^{t_i}$ is described by \eqref{def_bias_potential}.
\end{cor}

\begin{proof}
    We remark that if we replaced the family of intensities $\{\rho_{t_i}(\tilde m)\}_{\tilde m\in \mathbb N}$ by $\{\rho^{t_i}_{t_i}(\tilde m)\}_{\tilde m\in \mathbb N}$ in the definition of $\bar \rho_{t_i}$, the proof would be exactly the same as the one of Corollary \ref{cor_realm} using Proposition \ref{prop_caracteristion_rhoti} instead of Proposition \ref{prop_master_equation_realm}. 
    
    It is then enough to use Remark \ref{rem_uptoeps}, and the fact that $\sum\limits_{\tilde m \geq 0} \frac{1}{2^{\tilde m}} = 2$, to conclude.
\end{proof}

Corollary \ref{cor_conditionalm} allows us to partially understand the path-measure that would be reconstructed when applying the reweighting method described in Section \ref{subsec_MFL} to scRNA-seq datasets with lineage tracing when there is both death and subsampling. 

If the measurements are done at only two timepoints $t_1 = 0$ and $t_2 = T$, the reweighted empirical distributions correspond to $\hat \mu_0$ and $\hat p_T$ respectively, where $p_T$ is the solution of the PDE \eqref{SDE_withbias} characterizing a SDE with drift $-\nabla(\Psi + \Psi^{T})$ starting from $\mu_0$. In that case, as the drift of this SDE is a gradient, Theorem 2.1 of Lavenant et al.~\cite{Lavenant2021} holds and the reconstructed time-varying distribution obtained with the reweighting method is the one from the SDE with bias in Eq.~\eqref{SDE_withbias}, at least when $T \to 0$.

However, when the measurements are done at many timepoints $(t_i)_{i=1,\cdots,N}$, at each timepoint $t_i$ the reweighted distribution $\hat p_{t_i}$ is associated to the distribution $p_{t_i}$ of an SDE with timepoint-specific drift bias $-\nabla \Psi_{t_i}$. 
Thus, the time-varying distribution reconstructed by reweighting with observed generation numbers would be the temporal marginal of an SDE whose marginal corresponds, at any timepoint $t_i$, with the marginal of an SDE with drift $-\nabla(\Psi + \Psi^{t_i})$ (starting from $\mu_0$). This SDE can be defined, for example, between every pair of timepoints $(t_i, t_{i+1})$, as the solution of a Schr\"odinger problem with temporal marginals $p_{t_i}$ and $p_{t_{i+1}}$, but the associated drift cannot be expressed w.r.t the characteristics of the original process $\Psi, b$ and $d$. 
The solution to a Schr\"odinger problem between two timepoints necessarily has gradient drift, but whether the bias remains a gradient with more than two timepoints remains unclear.

The bias of this new drift, with respect to the ground-truth $-\nabla\Psi$, may well be bigger than $-\nabla \Psi^{t_{i+1}}$, since it must compensate for the fact that the drift bias of the SDE generating the initial condition $p_{t_i}$ is $-\nabla \Psi_{t_i}$ instead of $-\nabla \Psi^{t_{i+1}}$. An explicit characterization of this gradient drift with respect to $\Psi, b$ and $d$, in particular in the limit $N \to \infty$, remains an open question. 
\resolved{\af{I struggled with this paragraph. It's not clear exactly what it's trying to say about the SDE the reconstructed path law corresponds to. What does ``compatible'' mean?}\ev{Is it better ?} \af{It's better. I changed a couple more sentences and added the line at the end of the previous paragraph about when we know the drift is a gradient.}}

\bigskip

\begin{figure}
\includegraphics[trim=6cm 0cm 15cm 0cm, width=\textwidth]{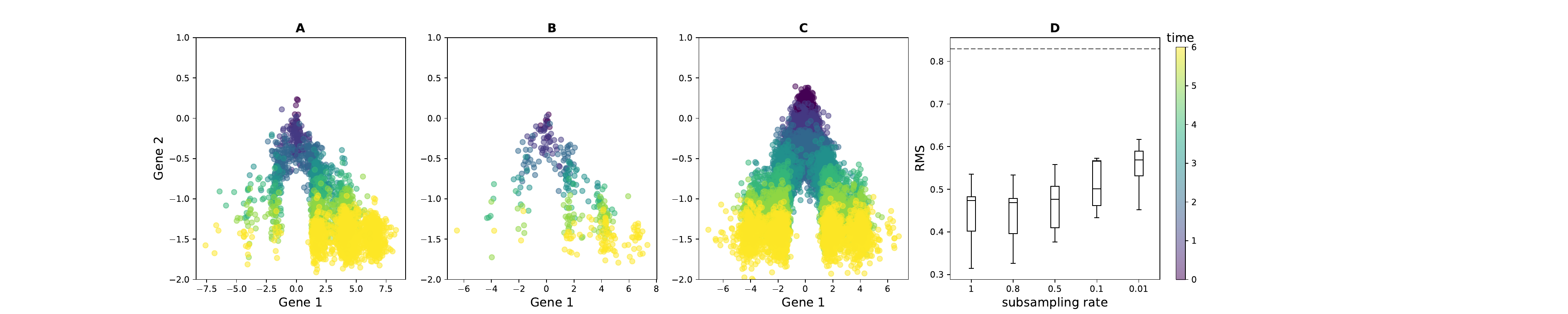}
\caption{Evolution of the bias w.r.t the subsampling rate. Simulations of the second branching SDE described in Appendix \ref{appendix_parameters} (A) without subsampling, (B) subsampling with a decreasing rate from 1 to 0.05, and (C) only the SDE without branching. In (D), we compare the cumulated RMS distance between the reweighted time-varying distributions and the ground-truth SDE at each timepoint for 5 different decreasing sequences of subsampling rates (including $q(t)=1$ where no subsampling occurs). Each box and whisker plot shows the distribution of RMS distances from 10 independent simulations. The subsampling rate indicated is $q(T)$, the rate for the last timepoint. The horizontal dashed line corresponds to the cumulated RMS distance between the simulations in (C) and the simulations of (A) without reweighting.
}
\label{Figure5}
\end{figure}

Our numerical experiments suggest the drift bias, while not fully characterized, will be small in practice. Fig.~\ref{Figure5} presents the evolution of the bias arising in the reconstruction of the time-varying distribution of the underlying SDE as the subsampling rate increases. Although, as expected, the quality of the reconstruction decreases with increasing subsampling rate, the loss in accuracy due to the subsampling is significantly smaller than the gain from reweighting. 
The cumulated RMS distance when reweighting after subsampling with $q(T) = 0.01$ (Fig.~\ref{Figure5}F, right) is 0.582, compared to 0.473 when reweighting with no subsampling (Fig.~\ref{Figure5}F, left) and 0.823 without reweighting (Fig.~\ref{Figure5}F, upper dashed line). The relatively small bias is not specific to the choice of the branching SDE's parameters: in our numerical experiments, we found it difficult to choose a potential, an associated branching rate, and a subsampling rate such that i) the trajectories show biologically meaningful structure, ii) Assumption \ref{assumption_survival} is satisfied, and iii) the bias is significant. In particular, we chose the more complex potential function of the two in Fig.~\ref{Figure2} for the simulations of Fig.~\ref{Figure5} because the effect of subsampling was not visible with the simpler two-well potential.

\bigskip

The main theoretical reason for this bias to be that small is that it is scaled by the diffusion coefficient $\tau$. In our simulations, even when significant differences exist in the branching rate between the potential wells, increasing $\tau$ causes the stochasticity from diffusion to erase the structure of the potential before it makes the drift bias $-\nabla \Psi^{t_i} (= \tau \nabla \ln S^{t_i})$ substantial.\resolved{\af{Does my rephrasing accurately describe what you saw, Elias?}\ev{Yes!}}  We discuss this relation further in Section \ref{subsec_estimation_subsampling_rate}.

While the bias may rarely be large, reducing it would improve inference accuracy. To do so, the next section introduces a heuristic method for removing the incompatibility between reweighted marginals by using more information from each tree $\mathcal T_i$ than the observable generation numbers we have restricted ourselves to thus far.

\subsection{Reducing the bias from death and subsampling}
\label{subsec_bias_reduction_method}
 
As described in the previous section, the reweighting method leaves a bias in the drift when cells are subsampled or die. While we believe removing the bias entirely is out of reach with our data (Remark~\ref{fundamental_limitation} below), we can nevertheless take steps to control it. We are now going to present a heuristic method for building, for each pair of timepoints $(t_i, t_{i+1})$, pairs of \emph{compatible} distributions $(\hat p^s_{t_i}, \hat p_{t_{i+1}})$ in the sense that they correspond to the temporal marginals of an underlying SDE whose drift is $-\nabla (\Psi + \Psi^{t_{i+1}})$. \resolved{\af{It's not clear what a bias ``$-\nabla \Psi^{t_{i+1}}$ between each pair of timepoint $[t_i, t_{i+1}]$" means.}\ev{Should be resolved}}

\bigskip 

The starting point of the method is that for any pair of timepoints $[t_i, t_{i+1}]$, we would like to interpret the family of intensities observed at $t_{i+1}$, $\{\rho_{t_{i+1}}(m)\}_{m \in \mathbb N}$, as the solution at time $t_{i+1} - t_i$ of a system of master equations of the form \eqref{system_master_equation_conditional} starting at $\{\rho_{t_i}(m)\}_{m \in \mathbb N}$. From proposition \ref{prop_caracteristion_rhoti}, that interpretation would be valid if we knew each cell alive at time $t_{i}$ would have least one descendant subsampled at $t_{i+1}$. However, in general it is quite possible for branches to die out between $t_i$ and $t_{i+1}$, in which case some cells observed at $t_i$ do not correspond to the ancestor of any cell at $t_{i+1}$.
If we were able to reconstruct a family of intensities $\{\rho_{t_i}^s(m)\}_{m \in \mathbb N}$ describing the cells sampled at time $t_i$ with at least one observed descendant at time $t_{i+1}$, the SDE described in Corollary \ref{cor_conditionalm} would connect this family of intensities to the intensities at $t_{i+1}$.

\bigskip

The new step in our bias-reduction method is reconstructing the family $\{\rho_{t_i}^s(m)\}_{m \in \mathbb N}$ from the observations. For this task, we can use more information about the lineage tree than the observable generation numbers that we have used until now. In particular, each lineage tree observed at time $t_{i+1}$ gives us the times of last common ancestor for every pair of cells.
Going backward in this lineage tree to find the structure at $t_{i}$ provides an intensity in $\mathcal M_+(\mathbb N)$ which coincides with $\rho_{t_i}^s(m, \mathcal X)$. 
Our algorithm makes an initial estimate of $\rho_{t_i}^s(m, \mathcal X)$ from the lineage tree $\mathcal T_{t_{i+1}}$ via a graphical model, in a similar fashion to the ancestor estimation step of LineageOT~\cite{forrow2021lineageot}. It then improves that estimate by combining it with $\{\rho_{t_i}(m)\}_{m \in \mathbb N}$ using partial optimal transport. 
The principles of the method are illustrated in Fig.~\ref{Figure7}, and fuller details on each step can be found in Appendix \ref{appendix_heuristic}.

\bigskip

Figure \ref{Figure6} presents the results of applying this bias reduction method to data obtained from the same potential as Fig.~\ref{Figure5}.
For each timepoint $t_i$, we compute the RMS distance between: i) the distribution obtained by simulating the SDE from the distribution at this timepoint (denoted $p_{t_i}^*$ for the standard MFL method, and $\mu_{t_i}^*$ for the algorithm presented in this section), and ii) the distribution computed by the method at the following timepoint (denoted $p_{t_{i+1}}^*$ for the standard MFL method, and $\nu_{t_i}^*$ for the algorithm presented in this section).
The result provided is the cumulated RMS distance on the timepoints. As expected, our method reduces the bias between each pair of timepoints. In particular, for the reweighted distributions (before applying the MFL algorithm), our method achieves a bias comparable with the the case without subsampling. For the distributions reconstructed with the MFL algorithm, our method still allows for a bias reduction. In that case, the difference between the case with and without subsampling is less important since the MFL algorithm itself, by smoothing the errors, already provides good results. This is in line with the simulations in Fig.~\ref{Figure5} which suggested that the bias induced by subsampling is small compared to the improvement reweighting provides.

 \begin{figure}
	    \includegraphics[width=1\textwidth]{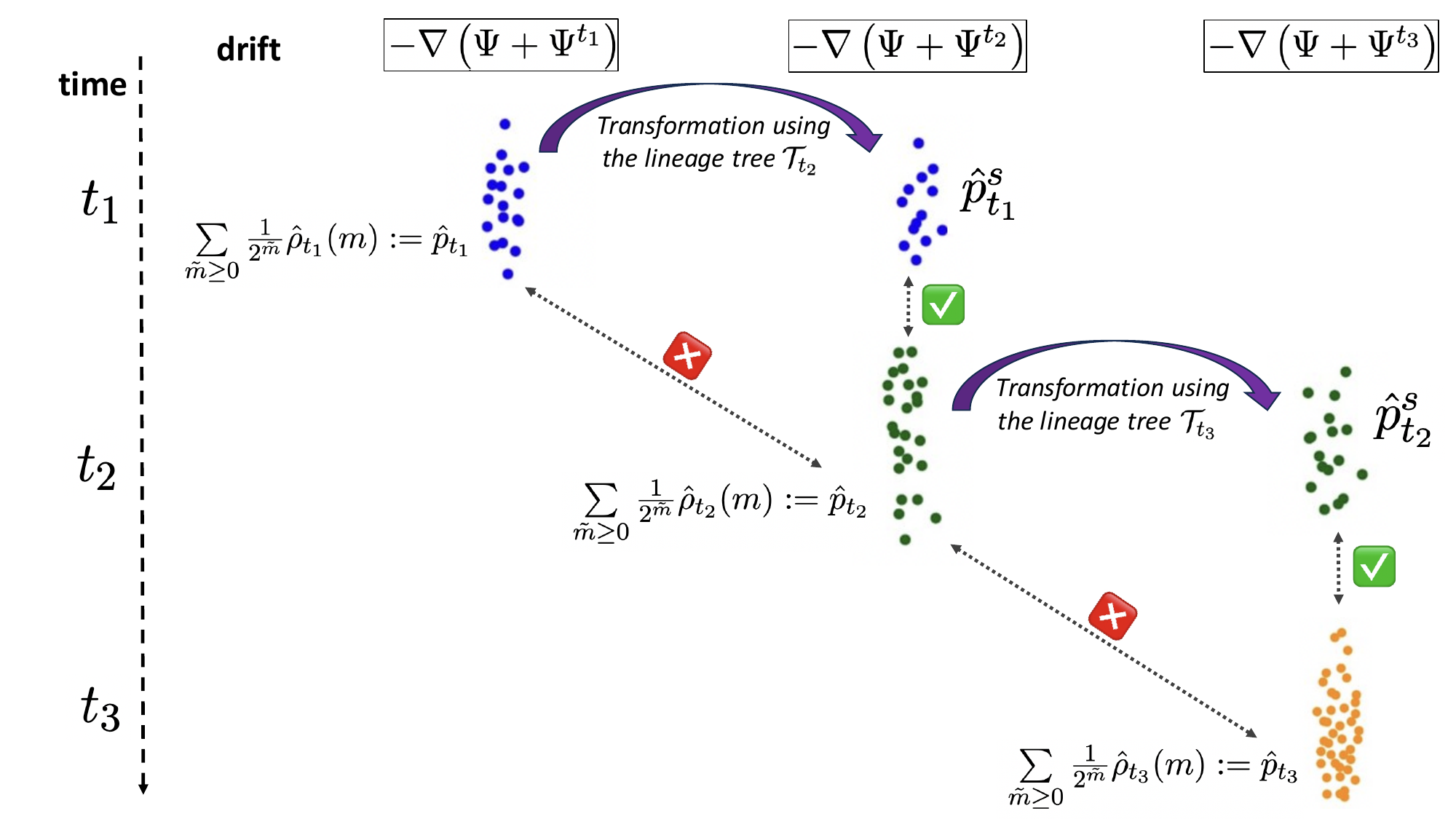}
     \caption{Representation of the framework described in Sections \ref{subsec_bias_reduction_method} to build for each pair of timepoints $(t_i, t_{i+1})$, pairs of \emph{compatible} distributions $(\hat p^s_{t_i}, \hat p_{t_{i+1}})$ in the sense that they correspond to the temporal marginals of an underlying SDE whose drift is known at the two timepoints (represented by a green square in the figure). First, we build at each timepoint $t_i$ the \emph{reweighted distribution} $\hat p_{t_i}$ from the observed family of intensities $\{\hat \rho_{t_i}(\tilde m)\}_{\tilde m \geq 0}$, corresponding to the temporal marginal of a SDE with timepoint-specific drift $-\nabla \left(\Psi + \Psi^{t_i}\right)$. As the pair $(\hat p_{t_i}, \hat p_{t_{i+1}})$ is not compatible (represented by a red square in the figure),  we then use the lineage tree $\mathcal T_{t_{i+1}}$ to approximate the distribution $\hat p^s_{t_i}$ that would have been obtained at $t_i$ with the drift $-\nabla \left(\Psi + \Psi^{t_{i+1}}\right)$. Note that without bias, that is with $\Psi^{t_i}=cste$ for all $i$, each pair $(\hat p_{t_i}, \hat p_{t_{i+1}})$ would be directly compatible and this scheme would consist only in a column instead of an array.} 
     \label{Figure7}
\end{figure}

\begin{figure}
\includegraphics[trim=6cm 0cm 15cm 2cm, width=\textwidth]{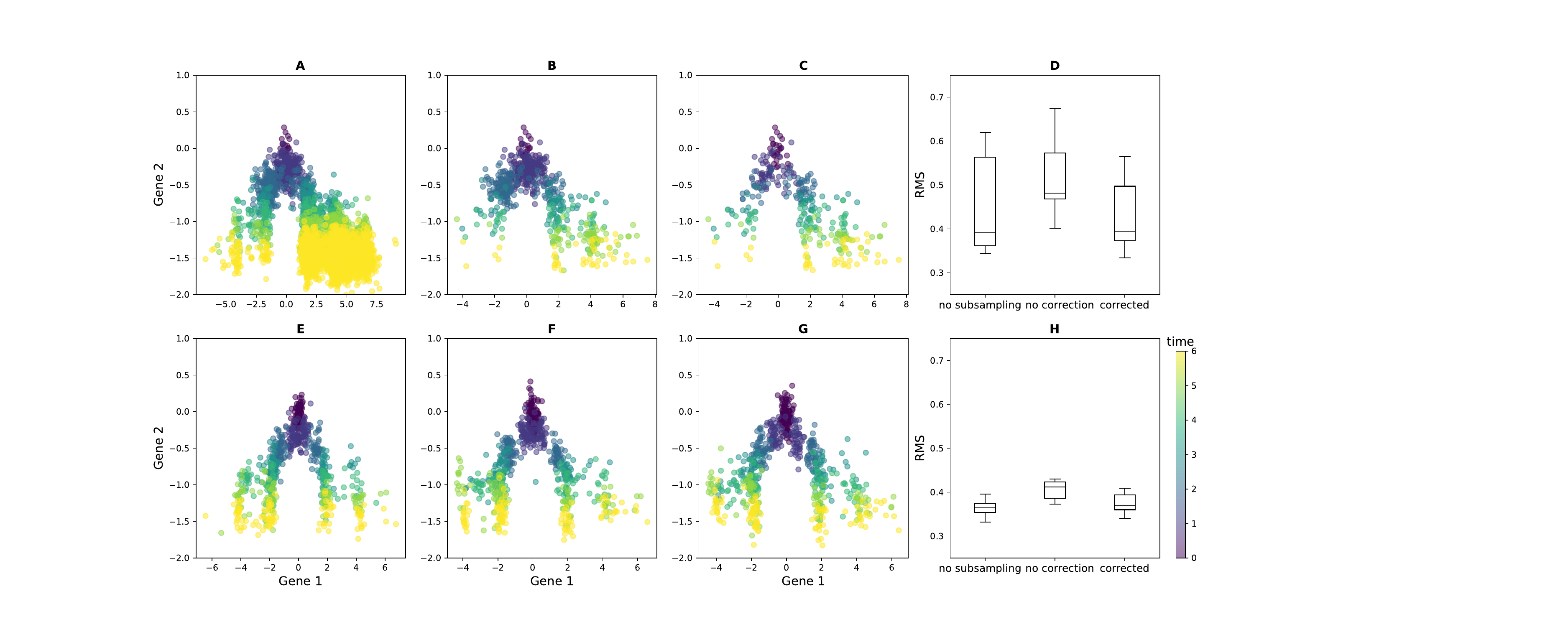}
\caption{Estimation of the method for reducing the bias described in Section \ref{sec_method_bias}. Simulations of the second branching SDE described in Appendix \ref{appendix_parameters} for (A) without subsampling, (B)-(C) subsampling with a decreasing rate from 1 to 0.05. In (C), the method described in Section \ref{subsubsec_step1} is applied to reconstruct pairs of timepoints that are compatible with the SDE \eqref{SDE_withbias}. (E)-(F)-{G} reconstruction with the MFL method of the trajectories of the SDE from the datasets of (A)-(B)-(C) respectively. We also represent for each cell the velocity field estimated with the method described in Section \ref{subsec_applications} (in black), compared to the real one (in red). In (D) and (H), we compare the cumulated RMS distance between the reconstructed empirical distributions at each timepoint and the one obtained by simulating cells from the empirical distribution reconstructed at the previous timepoint, for the datasets of (A)-(B)-(C) and (D)-(E)-(F) respectively.
\resolved{\af{The labels in this figure and the others are small enough to be hard to read.}\ev{Is it enough now?}}
}
\label{Figure6}
\end{figure}

\bigskip

\begin{rem}
\label{fundamental_limitation}
    The method presented in this section, in addition to losing the theoretical guarantees of the reweighting method, does not entirely remove the expected bias in the reconstructed drift. The results after correction in Fig.~\ref{Figure6}.F are not quite as good as the case without subsampling. The correction nevertheless has two advantages: the bias is better understood, and it is expected to become very small when the timepoints are close. We claim that the remaining bias reflects nonidentifiability in the data itself. Indeed, lineage tracing data are by nature unable to distinguish the cases in Figs.~\ref{Figure1}-B and \ref{Figure1}-C, and thus reconstructing the characteristics of an SDE without any bias appears out of reach without new experimental techniques.
\end{rem}

\subsection{Conditions for small bias}
\label{subsec_estimation_subsampling_rate}

In this section, we provide some quantitative and qualitative analysis aimed at estimating, from some minimal knowledge on the branching rates, necessary conditions on the subsampling probability and the death rate for the bias from death and subsampling to be small.

\bigskip

On one side, a consequence of Proposition \ref{prop_caracteristion_rhoti}, under Assumption \ref{assumption_survival} is that the marginal on $\mathcal X$ of the time-varying family of intensities associated to the branching SDE with observable generation numbers and subsampling, denoted $\tilde \rho$, corresponds at every timepoint $t_i$ to the intensity at $t_i$ of a new branching SDE without death, described by: 
$$\partial_t \rho^{t_i} = \mathcal L^{*}_{\Psi + \Psi^{t_i}, \tau}\rho^{t_i} + bS^{t_i} \rho^{t_i},$$
with the initial condition $\rho^{t_i}_0 = \mu_0$. 
If $\Psi^{t_i}$ were zero, this would be equivalent to the branching-free SDE (Eq.~\eqref{master_equation_SDE} with an additional birth term. As birth can only add mass, zero bias implies that for all $t,x$, $\tilde \rho_t(x) \geq p_t(x)$, where $p$ is the time-varying probabilistic distribution defined by \eqref{master_equation_SDE} with the same initial condition $\mu_0$. By substituting Eq.~\eqref{rel_intensity_subsampling} into this inequality, we have a necessary condition on the subsampling rate: 
\begin{align}
\label{relation_number_cells}
    \forall t \in [0,T],\, \forall x \in \mathcal X:\, q(t) \rho_t(x) \geq p_t(x).
\end{align}

On the other side, Corollary \ref{cor_conditionalm} allows us to partially understand what the reweighting method described in Section \ref{subsec_MFL} would give if applied directly to the case with death and subsampling. We expect the bias in the drift in that case to be bigger than $-\nabla(\Psi^{t_{i}})$ to compensate for the fact that each reweighted distribution $\bar \rho_{t_i}$ corresponds to SDE starting at $\mu_0$ with timepoint-specific drift $-\nabla(\Psi + \Psi^{t_i})$. Nevertheless, if $\Psi^{t_i}$ is small for all $t_i$, the total bias should likewise be small, as the incompatibility between marginals at different timepoints decreases with $\Psi^{t_i}$

\bigskip

We can combine Eq.~\eqref{relation_number_cells} with Corollary \ref{cor_conditionalm} to determine the subsampling probability $q(t)$ required for the reweighting method presented in Section \ref{subsec_MFL} to be accurate in the limit of small noise and nonnegligible branch survival probability $S^{t_i}$ (Eq,~\eqref{assumption_reg_survival}). The bias is scaled by $\tau$, while $\log(S^{t_i})$ remains $O(1)$ unless $S^{t_i}\to 0$. \resolved{\af{What difficulties arise if $S^{t_i}$ goes to zero and its log blows up? I suspect these limits work but require more care than is worth writing.}} Thus, as $\tau \to 0$, for every timepoint $t_i$ the intensity $\bar{\rho}_{t_i}$ converges to the marginals at $t_i$ of the ground-truth time-varying probability distribution $p$ defined by \eqref{master_equation_SDE}. The limiting factor is then Assumption~\ref{assumption_survival}, for which Eq.~\eqref{relation_number_cells} gives a necessary condition: on every trajectory solution of the dynamical system $\dot \phi = -\nabla \Psi(t, \phi)$, we must have
\begin{align}
    \label{relation_number_cells2}
    q(t)e^{\int_0^t (b-d)(s, \phi(s))\mathrm ds} \geq 1
\end{align}
 for all $t \in [0,T]$.
This last inequality can be interpreted as a way of using prior knowledge on the birth and death rates to estimate the minimum subsampling probability required for accurate results from the reweighting method. For example, taking $d = 0$, a rough condition on $q$ becomes:
$$q(t) \geq e^{- t\min\limits_{[0,t]\times \mathcal X} b}.$$
Similarly, if $d \neq 0$ and $q = 1$ uniformly on $[0,T]$, a rough condition for the relation \eqref{relation_number_cells2} to be satisfied is that $b - d \geq 0$ on $[0,T] \times \mathcal X$. 

\bigskip
We expect this analysis in the limit of small noise to remain relevant in the realistic situation where the diffusion coefficient $\tau$ is not close to $0$. Indeed, if the relation \eqref{relation_number_cells} is satisfied, the probability of a cell to have a descendant subsampled at each timepoint ought to be high everywhere, and then the bias defined by $\tau \nabla \ln S^{t_i}$ is close to $0$. Eq.~\eqref{relation_number_cells}, therefore, is both a necessary condition for the reweighting method to be accurate when $\tau\to 0$ and a heuristic quantitative condition for the bias to be reasonably small for any $\tau$.

\resolved{\begin{rem}
For some potentials and death rates, there will be initial distributions $\mu_0$ where Assumption \ref{assumption_survival} does not hold in the limit of small noise. Considering only cellular death and not subsampling, the event of a process going extinct can be strongly stochastic. For example, if some cellular states are associated to a high birth rate, in particular at the beginning of the process, as long as the probability of transitioning to these states is not too small the process is unlikely to go extinct. But without stochasticity, cells starting in other regions of the gene expression state may never reach these states of high proliferation. If these regions are not negligible for the initial distribution $\mu_0$, the process would go extinct before time $T$ with non-negligible probability.
\af{I'm not sure this remark is worth keeping. If we want it, could we move it up with Assumption~\ref{assumption_survival} so we end this section with the condition for small bias?}
\end{rem}
}

\section{Trajectory inference via division recording}
\label{sec_death_process}

We showed in Corollary \ref{cor_conditionalm} that the reweighted distribution obtained using the observable generation numbers $\tilde m(X)$ at each timepoint corresponds, under Assumption \ref{assumption_survival}, to the distribution of an SDE whose drift is biased with respect to the drift of the original SDE \eqref{master_equation_SDE}. On the other hand, when there is no death or subsampling, Corollary \ref{cor_realm} ensures that the distribution reweighted with the real generation numbers $m(X)$ corresponds to the marginal of the original SDE (Eq.~\eqref{master_equation_onlydeath} with $d=0$). Although $m(x)$ is not accessible with lineage tracing data, experimental techniques in development~\cite{masuyama2022molecular} suggest the possibility of recording the generation numbers directly.

That prospect naturally raises the question of whether observing $m(x)$ enables unbiased reconstruction of the drift and branching mechanism in situations with subsampling and death.
In this last section, we show that the answer to this question is positive, but only partially: while knowing $m(x)$ does allow extending the consistency result \eqref{thm_consistency_1} when there is subsampling but no death, it is not sufficient when the death rate is non-uniformly zero, in which case stronger assumptions and prior knowledge are required.

\bigskip

Let $P^{\tau,d}$ be the path-measure associated to a branching SDE with gradient-drift and only vanishing particles described by the equation \eqref{master_equation_onlydeath} with death rate $d(t, x)$, null birth rate, and initial condition $\mu_0$. Let $W^{\tau, d}$ be the law of a BBM with the same diffusivity $\tau$, death rate $d(t, x)$, null birth rate, and initial condition $\mu_0$. For any sequences $(t_i)_{i=0\cdots,T}$ and $\left (\hat \mu_{t_i} \right)_{i=0,\cdots,T}$, we define the new functional:
\begin{align}
\label{functional_FTLH_degradation}
    F_{N,\lambda,h}(\hat \mu):\, R\to\tau H(R|W^{\tau, d}) + \frac{1}{\lambda}\sum\limits_{i=1}^N ({t_{i+1} - {t_i}})H(\Phi_h * \hat \mu_{t_i} | R_{t_i}).
\end{align}
Finally, $\rho_t$ denotes the time-varying distribution of the ground-truth branching SDE with birth and death under which the data are sampled. The following result states that, provided that the death of cells can be neglected or that we know prior to the observations the true death rate, the minimizer of \eqref{functional_FTLH_degradation} is still consistent:

\begin{theo}
\label{thm_consistency_2}
Let $\hat \mu = \left (\hat \mu_{t_i} \right)_{i=0,\cdots,T}$, where for all $t_i$ $$\hat \mu_{t_i} = \frac{1}{K_i}\sum\limits_{j=1}^{k_i} \frac{1}{2^{m_{j,i}}}\delta_{x_{j,i}}$$
is the reweighted empirical distribution of tuples $\{(m_{j,i}, x_{j,i})\}_{j=1,\cdots,k_i}$ subsampled at rate $q(t_i)$ from $K_i$ trees. 
Let $R_{N,\lambda,h}^{d}$ be the minimizer of $F_{N,\lambda,h}(\hat \mu)$ among all path-measures of branching processes with initial mass equal to one and death rate differing from $d(t,x)$ by a function $C$ from $[0,T]$ to $\mathbb R$\resolved{\af{Elsewhere the notation $f:A\to B$ indicates $f(A)=B$; we should be consistent about whether we mean that or that $A$ and $B$ are the domain and codomain.}\ev{should be resolved}\af{I changed a few more. I don't see anywhere else left where it means domain/codomain.}} depending on $t$ but not $x$.
We have:
\begin{enumerate}
    \item If $d = 0$, in the limit $N \to \infty$, followed by $\lambda \to 0, h \to 0$, $R_{N,\lambda,h}^{d}$ converges narrowly to $P^{\tau}$ in $\mathcal P(\textrm{c\`adl\`ag}([0, T], \mathcal \mathcal X))$.
    \item If $d \neq 0$ and the subsampling probability function $q(\cdot)$ is decreasing on $[0, T]$, in the limit $N \to \infty$, followed by $\lambda \to 0, h \to 0$, $R_{N,\lambda,h}^{d}$ converges narrowly to $P^{\tau, \tilde d}$\\ in $\mathcal P(\textrm{c\`adl\`ag}([0, T], \mathcal M_+(\mathcal X)))$, where $\tilde d = d - \frac{\mathrm d}{\mathrm dt} \ln(q(t))$. \resolved{\nitya{Is this true? If $q(\cdot)$ is constant in $t$ then $\tilde{d} = d$.}\ev{You're right, I've precised that it is the case if we impose that the initial mass is $1$}}
\end{enumerate}
\end{theo}

\begin{proof}
The proof of the point 1. is similar to the one of Theorem \ref{thm_consistency_1}, but in this case the law of large numbers is applied to the renormalized time-varying intensity $q(t) \rho_t$. We obtain this time a weak convergence of the form:
\begin{align}
\lim_{T \to \infty} \sum\limits_i^{n} \omega_i^T \int_{\mathcal X}f(x)\hat \mu_{t_i}^h(m, \mathrm dx) =  \int_{\mathcal X}f(x)q(t)\sum\limits_{m=0}^\infty\frac{1}{2^m}\Phi_h*\rho_t(m, \mathrm dx).
\label{weak_convergence_TCL_2}
\end{align}
Theorems 4.4 and 4.6 of Lavenant et al.~\cite{Lavenant2021} then ensure that
$F_{N,\lambda,h}$ $\Gamma$-converges under Assumption \ref{assumption_initial_distrib} to $$F:\, R \to \mathds 1_{H(R|W^{\tau, 0}) < \infty} + \int_0^1 H\left(q(t_i)p_{t_i}\bigg| \mathbf E^R[Z_t]\right)\mathrm dt,$$
where $p_{t_i}$ denotes the probabilistic distribution characterizing the marginal at $t_i$ of the path measure $P^{\tau}$.
Point 1. follows, since only an SDE, which is a probabilistic process on $\mathcal X$, has finite entropy with respect to a Brownian motion. The minimum of $F$ among the probabilistic processes, where
$$F(R) = \int_0^1 H\left(q(t_i)p_{t_i}\bigg| R_t\right),$$
is precisely $P^{\tau}$.

\bigskip

The proof of point 2. begins with the observation that if $d \neq 0$, the $\Gamma$-limit of $F_{N,\lambda,h}$ becomes
$$F:\, R \to \mathds 1_{H(R|W^{\tau, d}) < \infty} + \int_0^1 H\left(q(t)\mathbf E^{P^{\tau, d}}[Z_t]\bigg| \mathbf E^R[Z_t]\right)\mathrm dt.$$
To conclude the proof, we claim that the minimizer of this problem, among all the branching SDEs for which there exists a function $C$ from $[0,T]$ to $\mathbb R$ such that the death rate is equal to $d + C$, is indeed $P^{\tau, \tilde d}$ (as soon as $P^{\tau, \tilde d}$ is well defined). 

It is easy to verify that if $\frac{\mathrm d}{\mathrm dt} q(t) \leq 0$ for every $t \in [0, T]$, $P^{\tau, \tilde d}$ is well defined and its intensity at $t$ is equal to 
$\exp\left(\int_0^t \frac{\mathrm d}{\mathrm ds} \ln(q(s))\mathrm ds \right)\mathbf E^{P^{\tau, d}}[Z_t] = \frac{q(t)}{q(0)}\mathbf E^{P^{\tau, d}}[Z_t]$. The fact that it minimizes the functional $F$ is justified by Lemma \ref{lem_consistency} presented below.
\end{proof}

\begin{lem}
\label{lem_consistency}
    With the notations of Theorem \ref{thm_consistency_2}, for all branching process\\
    $R^d \in \mathcal P(\textrm{c\`adl\`ag}([0, T], \mathcal M_+(\mathcal X)))$ such that $\mathbf E^{R^d}[Z_t] = \mathbf E^{P^{\tau, \tilde d}}[Z_t]$ for all $t \in [0, T]$, and with death rate equal to $d(t,x) + C(t)$ for some function $C$ from $[0,T]$ to $\mathbb R$, there holds
$$H(R^d\big | W^{\tau, d}) \geq H(P^{\tau, \tilde d}\big | W^{\tau, d}),$$
with equality if and only if $R^d = P^{\tau, \tilde d}$.
\end{lem}
\resolved{\af{Clarify notation that $d$ depends on $x$ and $t$; it's assumed known, not constant.}\ev{Should be resolved but to be careful throughout the manuscript...}}
\begin{proof}
Note that this lemma is a slight extension of Theorem 4.1 in~\cite{Lavenant2021}, and we refer to that article for further details. First, we restrict ourselves to processes $R^d$ such that $H(R^d | W^{\tau, d}) < \infty$, otherwise the result trivially holds. Denoting $p,r \in L^1(\textrm{c\`adl\`ag}([0, T], \mathcal M_+(\mathcal X)),W^{\tau, d})$ the Radon-Nikodym derivatives of $P^{\tau, \tilde d}$ and $R^d$, respectively, w.r.t $W^{\tau, d}$, we have by definition of the relative entropy \eqref{def_entropy} and by convexity of the function
$h:\, x\to x \ln x + 1 - x$ that, $W^{\tau, d}$-almost everywhere:
$$r \ln r − r + p - p \ln p \geq (r-p)\ln p,$$
with equality if and only if $r = p$. By integrating with respect to $W^{\tau, d}$, we obtain that:
$$H(R^d\big | W^{\tau,d}) - H(P^{\tau, d}\big | W^{\tau,d}) \geq \mathbf E^{R^d} [\ln p] − \mathbf E^{P^{\tau, \tilde d}}[\ln p].$$
Using the Ito formula for branching processes together with the form of the Radon-Nikodym derivative $p$ stated in~\cite{baradat2021regularized}, Theorem 4.19 and Theorem 4.23, respectively, and removing the terms that cancel on the right-hand side of the equation because they only depend on the marginal intensities of $P^{\tau, \tilde d}$ and $R^d$ , we obtain the inequality:
\begin{align*}
    H(R^d\big | W^{\tau,d}) - H(P^{\tau, d}\big | W^{\tau,d}) \geq \mathbf E^{R^d}\left(J\left[\ln\left(\frac{\tilde d}{d}\right) - \Psi\right]_T \right) - \mathbf E^{P^{\tau, \tilde d}}\left(J\left[\ln\left(\frac{\tilde d}{d}\right) - \Psi\right]_T \right),
\end{align*}
where $J[a]_t$ is a piecewise constant process which undergoes a jump of size $a(t, x)$ whenever a cell dies at time $t$ and position $x$ (see~\cite{baradat2021regularized}, Definition 4.15).

Because a process with finite entropy w.r.t. a branching SDE is a branching SDE (\cite{baradat2021regularized} Theorem 4.25), we can restrict to the case where $R^d$ is a branching SDE. It has a death rate $d^R = d + C$ by assumption, and any other branching mechanism has to be $0$ for this process to be of finite entropy with respect to $W^{\tau,d}$. From the proof of Proposition B.5 (\emph{ibid}), we also know that for any predictable field $a$ and any $t \in [0, T]$,
$J[a]_t - \int_0^t d^R \langle a(s), Z_s\rangle\mathrm ds$
is a martingale under $R^d$, and $J[a]_t - \int_0^t \tilde d \langle a(s), M_s\rangle\mathrm ds$ is a martingale under $P^{\tau, d}$. We then obtain:
\begin{align}
\label{inequality_lemma4.1}
    H(R^d\big | W^{\tau,d}) - H(P^{\tau, d}\big | W^{\tau,d}) &\geq \mathbf E^{R^d - P^{\tau, \tilde d}}\left( \int_0^T \left(d^R(t) - \tilde d(t)\right)\big\langle \ln\left(\frac{\tilde d(t)}{d(t)}\right) - \Psi(t), Z_t\big\rangle \mathrm dt\right),
\end{align}
and here $\rho_t$ denotes the common marginal intensity at $t$ of $P^{\tau, \tilde d}$ and $R^d$.

Now, as $P^{\tau, \tilde d}$ and $R^d$ are two branching SDEs with the same intensity, we may integrate their master equation \eqref{master_equation_BBM} on $\mathcal X$ for all $t$ to find:
$$\frac{\mathrm d}{\mathrm dt} \rho_t(\mathcal X) = \int_{\mathcal X} d^R(t,x) \rho_t(x)\mathrm dx = \int_{\mathcal X} \tilde d(t,x) \rho_t(x)\mathrm dx.$$
The difference $d^R(t,x) - \tilde d(t, x)$ therefore integrates to zero over $\mathcal{X}$. Because $d^R(t,x) - \tilde d(t, x)$ is independent of $x$ by assumption, we must have $d^R(t,x) - \tilde d(t, x) = 0$ (and hence $C(t) = -\frac{\mathrm d}{\mathrm dt} \ln(q(t))$). 

The quantity on the right-hand side of \eqref{inequality_lemma4.1} is thus equal to $0$. Equality happens only for $R^d =P^{\tau, \tilde d}$ by convexity of the relative entropy $H$, concluding the proof. 
\end{proof}

Theorem \ref{thm_consistency_2} is stronger than Theorem \ref{thm_consistency_1} in two ways: the convergence does not require the observation of all the cells of the observed trees, nor is it fundamentally limited to the case $d = 0$, as inference from lineage tracing data (see Remark \ref{fundamental_limitation}). However, it is also limited by three important factors:  i) it requires accurate prior knowledge of $d$, ii) there are no known algorithms for minimizing relative entropy while fixing a prescribed death rate, and iii) the required data are not yet directly accessible in the literature, even if in line with current development of measurement technologies~\cite{masuyama2022molecular}.

\section{Discussion}
\label{sec_discussion}

We have shown in this article that observing the generation numbers associated to the leaves of a tree characterizing a realization of a branching SDE is sufficient, in certain conditions, to deconvolve proliferation from the dynamics of individual cells. Our work extends the mathematical theory of trajectory inference developed for SDEs without proliferation to the branching case with similar theoretical guarantees and computational cost. Using the generation numbers, we are able to not only accurately estimate the drift of the SDE, as has been done for the case without branching, but also learn the proliferation rate with no prior knowledge. In particular, in the limit of infinitely close timepoints, we can rigorously reconstruct the law of the SDE modeling the motion of cells in gene expression space from time-series of measures from the process with proliferation, if:
\begin{enumerate}
    \item We have access to the \emph{observable generation numbers} and we observe all the leaves of at least one full tree per timepoint and there is no death (Theorem \ref{thm_consistency_1});
    \item We have access to the \emph{real generation numbers} and we observe cells that are subsampled uniformly on at least one tree per timepoint and: i) there is no death or ii) the subsampling rate is known or decreasing, and the death rate is known prior to the observations (Theorem \ref{thm_consistency_2}).
\end{enumerate}

These results demonstrate that much of the information in the lineage trees is not needed for reconstructing the trajectories of branching processes when there is no death nor subsampling. They do not use the times of last common ancestors between any pair of cells and the joint distribution of leaves. If only the observable generation numbers are available and cells are subsampled and can die, we have shown that the method for case 1 generates a bias in the reconstructed drift and branching rate. This bias is expected to be small in biologically relevant scenarios where the probability of having at least one observed descendant does not vary dramatically across cell. It can also be partially removed using additional information from the lineage tree via the algorithm of Section~\ref{subsec_bias_reduction_method}, albeit without the theoretical guarantees proved for the two cases above.

\bigskip

The main property underlying our results, that the reweighted marginal on $\mathcal X$ of a joint distribution on $\mathbb N \times X$ of a branching SDE corresponds to the probability distribution of the SDE without branching (see Corollary \ref{cor_realm}), with a possible bias due to death (see Corollary \ref{cor_conditionalm}), holds for general branching processes. Although we focused in this article on branching SDEs, these properties could be directly applied to account for branching in models with other experimentally-relevant stochastic processes, such as the gene regulatory network inference method based on switching ODEs we recently introduced~\cite{ventre2023one}.

\bigskip

Altogether, we believe that these results present a major theoretical advance in the nascent field of single-cell data analysis with lineage tracing. We go beyond the point of view developed in methods previously published to analyze these data~\cite{forrow2021lineageot,wang2022cospar,lange2023mapping}, which were only interested in reconstructing couplings between empirical measures of cells. As the notion of coupling between cells of empirical non-probabilistic measures is itself hard to define properly, it seems difficult to relate the results provided by these methods to the notion of path-measure of an underlying branching process, which is the aim of trajectory inference. To tackle this problem, our method reconstructs well defined couplings between cells belonging to the temporal marginals of the SDE underlying the observed branching SDE: considering these probabilistic distributions instead of the non-probabilistic observable measures is one of the major strengths of our work. Moreover, we emphasize that this underlying SDE is the main mathematical object of interest for biologists, as it shapes the called Waddington landscape~\cite{waddington2014strategy, huang2005cell} of differentiation, which encodes the regulatory mechanisms controlling cell fates. In reconstructing the law of this SDE from scRNA-seq measurements with lineage tracing, our method extracts the most important information from the data. Combined with estimation of the branching rate, we achieve a full understanding of the branching SDE when there is no death.

\section*{Code availability}

The code for reproducing the figures of this article is available at \\\url{https://github.com/eliasventre/LTreweighting}.

\section*{Acknowledgments}

GS and OA were supported by a NSERC Discovery Grant, and GS was supported by a MSHR Scholar Award.
We would like to thank especially Aymeric Baradat for his help with the theory of the unbalanced Schr\"odinger problem and enlightening discussions, and Stephen Zhang for help with the code of the MFL algorithm used in Figs. \ref{Figure3}, \ref{Figure4} and \ref{Figure6}.

\printbibliography

\begin{appendix}
\addtocontents{toc}{\setcounter{tocdepth}{-1}}
\section{Parameters of the simulations}
\label{appendix_parameters}

The applications of this article focus on two very simple potentials defined in $\mathbb R^3$, which are both associated to non-constant birth rates. For both potentials, the branching rate is built such that it is constant in every potential well: this is in line with the interpretation of the potential wells as cell types, and the fact to consider that the stochasticity is characterized by the transitions between these cell types~\cite{ventre2021reduction}.

\bigskip

We define $A = (1.5, 0, 0), B = (-1.5, 0, 0)$. The first potential, corresponding to the simulations of the first row of Fig.~\ref{Figure2} and of the simulations of Figures \ref{Figure3} and \ref{Figure4} is the following:
$$\forall x = (x_0, x_1, x_2) \in \mathbb R^3:\,V_1(x,t) = (x_0 - A)^2 (x_0 - B)^2 + 10 (x_1 + t)^2 + 10 (x_2)^2.$$
The birth rate is:
$$b_1(x, t) = 3(\tanh(2x_0) + 1).$$
It has two wells at any timepoint.

The second one, corresponding to the simulations of the second row of Fig.~\ref{Figure2} and of the simulations of Figures \ref{Figure5} and \ref{Figure6} is the following:
$$\forall x = (x_0, x_1, x_2) \in \mathbb R^3:\,V_2(x,t) = 10*(\cos(x_0) - \textrm{abs}(x_0)) + 10 (x_1 + t)^2 + 10 (x_2)^2.$$
The birth rate is:
\begin{align*}
    b_2(x, t) = \begin{cases}
    2(\tanh(x_0/5) + 1) &\textrm{ if } x_1 \leq -0.5 \textrm{ and } x_0 \leq 6,\\
    4(\tanh(x_0/5) + 1) &\textrm{ if } x_1 \leq -0.5 \textrm{ and } x_0 > 6,\\
    10 &\textrm{ if } x_1 > -0.5.
\end{cases}
\end{align*}
It has an infinite number of wells, but at the time of simulations the SDE has only explored between 4 and 6 wells at the last timepoint (see Fig.~\ref{Figure2}).
    
\bigskip 

For the sake of simplicity, we consider the death rates $d_1 = d_2 = 0$ on $\mathbb R^3$.

\section{Algorithm for the heuristic method described in Section \ref{sec_method_bias}}
\label{appendix_heuristic}

We present here the details of the two-step algorithm aiming to use the lineage information to reduce the bias arising in the reweighting method in case of death and/or subsampling. The results obtained with this method are illustrated in Fig.~\ref{Figure6}.

\bigskip

The principle of this heuristic method relies on an additional assumption: we consider that for every cell subsampled at a timepoint $t_{i+1}$, an ancestor of it would have been subsampled at the previous timepoint $t_i$ with high probability. The underlying idea is that the total mass of the process is expected to increase in $[0,T]$ and that the subsampling probability $q(t)$ is decreasing (\emph{i.e} there are more and more cells in the organism but we subsample a smaller and smaller fraction of them). Thus, if the intensity of an area $A_{i+1} \subset \mathcal X$ is high enough at time $t_{i+1}$ for a cell to be subsampled, the ancestors of the cells in this area should be mostly concentrated at time $t_i$ in an area $A_i$, such that the probability of subsampling a cell there should be high too.

Note that if the subsampling probability is uniformly $1$ in $[0,T]$, so that the condition for a cell to be subsampled is only that it is alive, this hypothesis only means that every cell alive at a certain time has an ancestor alive at earlier timepoints, which is obviously verified.

\subsection{Reconstruction of the family of intensities $\{\rho_{t_i}^s(m)\}_{m \in \mathbb N}$}
\label{subsubsec_step1}

In practice, we only have access at each timepoint to a sum of $n$ diracs on $\mathbb N \times \mathcal X$ reweighted by the number of trees $K$, of the form $\frac{1}{K} \sum_{c = 1}^n \delta_{(m_c, x_c)}$. For the sake of simplicity, we consider the family of observations as empirical intensities on $\mathbb N \times \mathcal X$. From these observations $\hat \rho^s_{t_i}(\tilde m, \mathcal X)$, $\hat \rho_{t_i}$ and $\hat \rho_{t_{i+1}}$, we propose thus to estimate $\hat \rho_{t_i}^s$ using the following three-steps algorithm:
\begin{enumerate}
    \item Use a graphical model to estimate the positions of the cells in $\hat \rho^s_{t_i}$ from the cells in $\hat \rho_{t_{i+1}}$ using the lineage tree, with a method similar to the first step of LineageOT~\cite{forrow2021lineageot}. We obtain a first estimation $\hat \rho^{s1}_{t_i}$;
    \item Minimizing a partial optimal transport problem between $\hat \rho^s_{t_i}$ and $\hat \rho_{t_i}$ with a cost defined between tuples in $\mathbb N \times \mathcal X$. We obtain an optimal transport coupling $T$ between $\hat \rho_{t_i}$ and $\hat \rho^{s1}_{t_i}$;
    \item For every cell $(m, x) \in \hat \rho^{s1}_{t_i}$, substitute:
    $$x_{new} = \sum_{y \in \rho_{t_i}} y T(y, x).$$
    $\hat \rho^s_{t_i}$ is defined as the empirical measure defined on this new positions $x_{new}$ together with the same generation numbers.
\end{enumerate}

\subsection{Trajectory inference for a SDE with a known drift with reduced bias}
\label{subsubsec_step2}

With this new sequence of empirical intensities $(\hat \rho^s_{t_i})_{i=1,\cdots,T}$ in hands, we can then build a minimization problem allowing to estimate, as in Section \ref{subsec_applications}, the drift and the branching mechanism of the branching SDE. For this, we aim to extend the MFL algorithm used in Section \ref{subsec_applications} for 2 sequences of empirical intensities: the observed sequence $(\hat \rho_{t_i})_{i=1,\cdots,T}$ and the reconstructed one $(\hat \rho^s_{t_i})_{i=1,\cdots,T-1}$. We recall that in a limit of an infinite number of trees, each $(i+1)^{th}$ intensity in the first sequence is expected to represent the evolution, under the law of the process defined by \eqref{SDE_withbias}, of the $i^{th}$ intensity of the second sequence.
Thus, in this limit, minimizing the Schr\"odinger problem between each of these pairs would provide the coupling of an SDE with a gradient drift, the law of which is compatible with the one of the SDE defined by \eqref{SDE_withbias}. 
\bigskip

We first consider their associated reweighted empirical distributions, that we denote  $\hat{\bar \rho}_{t_i}$ and $\hat{\bar \rho}^s_{t_i}$, respectively. We would like, as in Chizat et al.~\cite{zhang2022trajectory}, to tackle the problem of small samplings by optimizing these new sequence of reweighted intensities.

For this reason we decide, in addition to minimizing a Schr\"odinger problen between each  pair of reweighted distributions ($\hat{\bar \rho}^s_{t_i}$, $\hat{\bar \rho}_{t_{i+1}}$), to chain all these pairs together by minimizing an unbalanced Wasserstein distance between $\hat{\bar \rho}_{t_i}$ and $\hat{\bar \rho}^s_{t_i}$, as by construction $\hat{\bar \rho}^s_{t_i}$ is expected to be close to $\hat{\bar \rho}_{t_i}$, but not equal.

We thus find two new sequences of empirical distributions $\mu_{t_1},\cdots,\mu_{t_N}$ and $ \nu_{t_1},\cdots,\nu_{t_N}$, each being described by an empirical measure with $M$ diracs, solving the following problem:
\begin{align}
    \label{dynamical_MFL_new}
    \inf\limits_{\mu, \nu}\left\{G(\mu,\nu) + \tau F(\mu,\nu) \right\},
\end{align}
with $$G(\mu,\nu) = \sum\limits_{i = 1}^{N-1} T_{\tau_i}(\mu_{t_i}, \nu_{t_{i}}) + \sum\limits_{i = 1}^{N-2} OT(\nu_{t_{i}}, \mu_{t_{i + 1}}) +  \frac{1}{\lambda}\sum_{i=1}^{N - 1} \Delta t_i\left[H\left(\hat \rho_{t_i + 1} | \Phi_h * \nu_{t_i}\right) +  H\left(\hat \rho^s_{t_i} | \Phi_h * \mu_{t_i}\right)\right],$$
and $F(\mu,\nu) = \sum\limits_{i = 1}^{N-1} H(\mu_{t_i}) + H(\nu_{t_{i+1}})$.

As in Chizat et al.~\cite{zhang2022trajectory}, the functional to minimize is convex in the $\mu_{t_i}$ and $\nu_{t_i}$ separately, and we can use a similar algorithm to solve the minimization problem and find the optimal empirical probabilistic distributions corresponding to the optimal pairs $(\mu_{t_i}^*, \nu_{t_i}^*)$. The code of these algorithms is available on Github.
\end{appendix}
\end{document}